\documentclass[a4paper,fleqn,usenatbib]{mnras}
\usepackage{graphicx}	
\usepackage{amsmath}	
\usepackage{amssymb}	
\usepackage{comment}
\usepackage{orcidlink}
\usepackage{threeparttable}
\usepackage{caption}
\usepackage{graphicx}
\usepackage{subcaption}

\def\lsun{{\rm L_{\odot}}}
\def\msun{{\rm M_{\odot}}}
\newcommand\rsun{{\rm R_{\odot}}}

\def\be{\begin{equation}}
\def\ee{\end{equation}}

\def\del#1{{}}

\newcommand\mearth{{\,{\rm M}_{\oplus}}}
\newcommand\mj{{\,{\rm M}_{\rm J}}}
\newcommand\rj{{\,{\rm R}_{\rm J}}}

\newcommand\MSunPerYear{~${\rm M_{\odot}}$~yr$^{-1}$\,}

\title[Disc Fragmentation: FFP ejections]{Disc fragmentation. II. Ejection of low mass Free Floating Planets from growing binary systems}

\author[Nayakshin et al.]{Sergei Nayakshin$^{1 \orcidlink{0000-0002-6166-2206}}$\thanks{sergei.nayakshin@le.ac.uk}, Luyao Zhang$^{1}$, Aleksandra \'Calovi\'c$^1$, Hans Lee$^{1}$, 
\newauthor 
Clement Baruteau$^2$, Farzana Meru$^{3,4}$, and Lucio Mayer$^5$\\
$^{1}$School of Physics and Astronomy, University of
  Leicester, Leicester, LE1 7RH, UK. \\
$^2$ IRAP, Universit\'e de Toulouse, CNRS, Universit\'e Paul Sabatier, CNES, Toulouse, France\\
$^3$ Centre for Exoplanets and Habitability, University of Warwick, Coventry CV4 7AL, UK \\
$^4$ Department of Physics, University of Warwick, Coventry CV4 7AL, UK\\
$^5$ Department of Astrophysics, University of Zurich, Winterthurerstrasse 190, 8057, Zurich, Switzerland\\
}

\date{Accepted XXX. Received YYY; in original form ZZZ}

\pubyear{2025}

\begin{document}
\label{firstpage}
\pagerange{\pageref{firstpage}--\pageref{lastpage}}
\maketitle

\begin{abstract}

Observations indicate that disc fragmentation due to Gravitational Instability (GI) is the likely origin of massive companions to stars, such as giant planets orbiting M-dwarf stars, Brown Dwarf (BD) companions to FGK stars, and  binary stars with separations smaller than $\sim 100$~au. Additionally, we have recently showed that disc fragmentation in young rapidly evolving binary systems ejects an abundant population of massive Jupiter-mass Free-Floating Planets (FFPs). In this model, a massive disc around an initially single protostar undergoes GI and hatches a number of fragments; the most massive oligarch grows by runaway accretion into the secondary star. As the system rearranges itself from a single to a binary star configuration, a dramatic "pincer movement" by the binary ejects planets through dynamical interactions with the stars. Here we propose that the same scenario applies to an even more abundant population of smaller FFPs discovered by the microlensing surveys. Although disc fragmentation is usually believed to form only massive objects, several pathways for forming small core-dominated planets at separations of tens of au exist. We present results from three complementary simulation approaches, all of which confirm  planet ejection efficiency as high as $\sim 50$\% for secondaries more massive than $\sim 10$\% of the primary star mass. On the other hand, Jovian mass planets migrate through the region of tens of au too rapidly to eject planets from that region. We discuss implications of this scenario, concluding that microlensing FFPs  may be the most convincing evidence yet that disc fragmentation forms low mass ($M_{\rm p}\ll 1\mj$) planets.
\end{abstract}


\section{Introduction}

Observations find that there are at least $\sim 2$ planets with radius $R_{\rm p} < 4 R_\oplus$ per M-dwarf star \citep{Dressing_13_planet_occrence,Clanton_16_Planet_mass_function,Mignon_25_planet_occurence}, the most common star in the Galaxy \citep{ChabrierBaraffe2000-Review}. Due to observational biases, these planets orbit their stars mainly at $\lesssim 1$ au. The Core Accretion theory for planet formation \citep[CA;][]{PollackEtal96,IdaLin04a,AlibertEtal05} provides convincing explanations for a significant number of statistical correlations found  in the data \citep{IdaLin04b,Mordasini14,Bern20-2}.

On the other hand, CA models predict very few gas giant planets around M-dwarfs \citep[][]{IdaLin_05_Low_Mstar,Liu_19_CA_low_Mstar,Burn_21_BernCA_lowMstar}, yet such planets are observed \citep[e.g.,][]{Bryant_25_giant_around_Mdwarf}. For example, CA population synthesis by \cite{Liu_19_CA_low_Mstar,Miguel_20_CA_Mdwarfs,Mulders_21_CA_Mdwarfs} forms no gas giants for stars with $M_* \leq 0.3\msun$.  Therefore, planet formation by disc fragmentation \citep[also often called Gravitational Instability scenario, GI;][]{Boss97,HelledEtal13a,KratterL16} is called for  \citep[e.g.,][]{Schlecker_22_giants_in_Mdwarfs,Bryant_25_giant_around_Mdwarf}. Although disc fragmentation is understood to occur at large separations from the star only \citep[$R\gtrsim$ tens of au; e.g.,][]{Rafikov05,Clarke09,KratterEtal10}, numerical simulations show that GI planets migrate inward very rapidly \citep{BoleyEtal10,VB10,MachidaEtal11,MichaelEtal11,BaruteauEtal11}. Support for this scenario comes \citep[][]{VB10,Nayakshin-23-FUOR,Nayakshin23-FUORi-2} from FU Ori outbursts of young protostars \citep[e.g.,][]{Fischer-PPVII}, many of which are M-dwarf stars \citep[see][]{Nayakshin-24-classic-TI}.

Effective at separations $R\gtrsim 2$~au, the technique of gravitational microlensing \citep[e.g.,][]{Mao_Paczynski_91,Gould_Loeb_92,Mroz-23-Microlensing-planets-review} finds that there is $\sim 0.75$ planet per microlensing host star \citep[e.g.,][]{SuzukiEtal16,Zang_25_bound_microlensing}. 
\cite{SuzukiEtal18} found that CA models under-predict the abundance of microlensing planets with masses $M\gtrsim 10\mearth$ by about an order of magnitude. Similarly, the 20 year-long HARPS radial velocity survey of M-dwarf stars finds \citep{Mignon_25_planet_occurence} that the disagreement between CA models and their observations is most significant for planets with masses $M_{\rm p}> 30\mearth$ in orbits with separations greater than $\sim 2$~au (cf. their Fig. 10). It is possible that, similarly to gas giant planets in shorter orbits around M-dwarfs, these planets also formed by disc fragmentation; their migration has simply terminated sooner \citep[e.g.,][]{NayakshinFletcher15}.

Formation of most massive (in terms of $q$, object mass to $M_*$ ratio) sub-stellar companions via disc fragmentation is also supported by the observed correlations of their frequency of appearance, $f_{\rm ap}$,  with stellar metallicity, [Fe/H]. In FGK hosts, $f_{\rm ap}$ does not correlate with [Fe/H] for planets more massive than $M_{\rm p}\sim 5\mj$ and BDs \citep[e.g.,][]{TroupEtal16,Schlaufman18}. This is contrary to predictions of the CA scenario \citep[][]{MordasiniEtal12}, but is consistent with predictions of disc fragmentation \citep[e.g., \S 9 in][]{Nayakshin_Review}. Further, the binary fraction of stellar primaries that host massive gas giants and BDs is $\sim 3$ times higher than that of the stars in the field \citep{Fontanive_19_giants_in_wide_binaries}. This is relevant because the dominant population of binary systems (those closer than $a_{\rm b}\lesssim 200$~au) is believed to also form by disc fragmentation (\S \ref{sec:binary_statistics}).

The microlensing Free-Floating Planets ($\mu$FFPs) are planets where no stellar host has so far been detected; these can be truly unbound planets or planets on very wide, $R \gtrsim 10$~au, orbits \citep{Mroz_17_Microlensing,Gould-22-microlensingFFPs,sumi2023}. \cite{Mroz-24-evidence-for-FFPs} followed up five microlensing event fields with deep Keck telescope observations and found no evidence for the presence of a stellar mass host in any of these targets, but they emphasize that deeper observations are needed to confirm this. This population is surprisingly numerous, with \cite{sumi2023} quoting $22^{+22}_{-13}$ $\mu$FFPs with mass $M_{\rm p} > 0.3 \mearth$ per microlensing host star, and \cite{Mroz-23-Microlensing-planets-review} noting that three major studies with three different sources of data yield consistent results, $7^{+7}_{-5}$ $\mu$FFPs with mass $M_{\rm p} > 1 \mearth$ per star, and the mass function dominated by low mass objects in terms of planet number, $dN/dM_{\rm p}\propto M_{\rm p}^{-2}$. Interestingly, \cite{Gould-22-microlensingFFPs} points out that this mass function, if extrapolated $\sim 17$ orders of magnitude in mass down, predicts the right number of ’Oumuamua type unbound asteroid-like objects \citep{Mamajek_17_Oumuamua}.

CA scenarios do not eject enough planets to account for the observed $\mu$FFP population. An efficient ejection of planets via N-body dynamical instabilities \citep{Weidenschilling_96_FFPs,Rasio_96_FFPs,juric2008} requires many $\sim$ Jupiter-mass planets per system, and it also leaves $1-2$ bound gas giant planets behind \citep{Chatterjee_08_PP_scattering}. However, the fraction of FGK stars with a gas giant at any separation is $\lesssim 0.25$ \citep{Fulton_21_giants_occurence}, far too low for this channel to work in practice \citep{veras-12-FFPs}. Furthermore, in M-dwarfs this fraction drops to just $\sim 0.06$ \citep{Clanton_16_Planet_mass_function,Zang_25_bound_microlensing}. Recognising this, \cite{Hadden_25_Neptune_FFP} consider systems of 5 Neptune mass planets, showing that many of these are ejected by N-body interactions. However, (i) CA planet formation models \citep{Burn_21_BernCA_lowMstar} produce $\lesssim 1$ Neptune per M-dwarf star \citep[cf. Fig. 10 in][]{Mignon_25_planet_occurence}; (ii) it is also not clear how the Neptune mass planets scenario would account for the more massive FFPs -- it is typically the less massive planets that are ejected; (iii) Neptune mass planets should grow by rapid gas accretion into gas giant planets \citep[e.g.,][]{Stevenson82,IkomaEtal00,IdaLin04a,MordasiniEtal12a,Piso15}, so this scenario likely over-produces the observed number of giants if the planet accretion physics is taken into account.

Binary systems can be very effective in ejecting planets in unstable orbits \citep{Holman_Wiegert_99}. However, forming planets via CA in such orbits is very challenging: the much more dynamic environment of binary stars is hostile to planetesimal growth \citep{PaardekooperEtal12,Marzari_13_CB_planet_formation,LinesEtal14,Marzari_19_CB_planets}. Due to this, planet formation physics constrains the types of systems in which planets can form in the CA scenario \citep{Thebault_15_Stype_binaries}. \cite{PierensNelson08} found that planets formed in circumbinary discs of close separation binaries $a_{\rm b}\lesssim 1$~au at ``safe formation" distances \citep[e.g.,][]{LinesEtal14} of $\sim 5$ or more binary separations, can be ejected by close scatterings with the secondary if they migrate sufficiently close to it \citep[see also][]{Coleman-23-planet-in-CB-systems}.  Despite such binaries representing only a small fraction of the overall population, \cite{Coleman-24-FFPs} find that they may yield sufficient numbers of ejections to account for FFPs with masses $M_{\rm p}\sim 10\mearth$. However, the scenario yields too few planets both below and above this mass \citep{Coleman-24-FFPs}, see also \S \ref{sec:CA_FFP} below. 

Here we propose that the populations of $\mu$FFPs, gas giants in M-dwarfs, and very massive gas giants and BDs in FGK hosts, are all related by their formation pathways. We hypothesize that FFPs form by fragmentation of protoplanetary discs around {\em initially} single Class 0/I protostars at separations of tens to $\sim 100$ au, but then find themselves in a growing binary system. In our model, in single stars, disc fragmentation results in substellar mass objects only. In FFP progenitor systems, the discs become particularly  massive and extended for a {\em seed} of the secondary star to form at distances $R\gtrsim 100$~au. The ensuing runaway accretion growth of the secondary star, plus the rapid shrinking of its orbit, leads to an extremely violent rearrangement of the system. This process places many of the planets in the system in unstable configurations naturally, and they are readily ejected.

These ideas were already explored by \cite{Calovic_25_FFP-1}, who used 3D hydrodynamical simulations to study evolution of $M_{\rm p}=(1-3)\mj$ gas giant planets and a seed of a growing secondary star embedded in a massive self-gravitating disc. The planets in the study were injected into the disc at radii of tens to $\sim 100$~au, whereas the secondary object was inserted on a much larger orbit, at $R= 140$~au. Abundant close planet-secondary interactions take place, and as much as $70\%$ of the planets were ejected in less than 0.1 Myr in the simulations.

In this paper we extend the scenario on all types of $\mu$FFPs. While disc fragmentation is usually credited with formation of just gas giant planets, BDs and binary systems \citep{KratterL16},  three ways of producing low mass planets, both gaseous and solid core-dominated, exist:

\begin{enumerate}
    \item 
    Dust in self-gravitating discs can be trapped in spiral density arms efficiently \citep{RiceEtal04,BoleyDurisen10,GibbonsEtal12,Shi_16_GI_dust_focusing,BoothClarke16}. Many authors find that these gas-enabled dust concentrations can collapse gravitationally into solid bodies as massive as $\sim 10\mearth$ \citep{GibbonsEtal14,Baehr_22_GI_dust,Baehr-23-GI-dust,Longarini_23_solids_collapse_theory,Longarini_23_solids_collapse_sims}.
    \item Another possibility is that gas disc does fragment onto self-gravitating clumps. Within the clumps, grains grow and sediment to the centre \citep{Kuiper51b,HW75,Boss98,HS08}, forming massive solid cores. The cores are then released back into the disc when the gas clump (core's envelope) is tidally disrupted \citep[e.g.,][]{BoleyEtal10,ChaNayakshin11a}. 
    \item \cite{Deng20-MHD-GI,Deng_21_GI_Neptunes,Kubli-GI-MHD-23} show that self-consistent inclusion of magnetic fields into self-gravitating disc simulations may yield bound gas clumps with mass as small as that of Neptune. 
\end{enumerate}


\section{Preliminaries: observations and past modelling}\label{sec:background}

\subsection{Binary statistics and formation}\label{sec:binary_statistics}

The Multiplicity Fraction is $\sim 25$\% for M-dwarfs \citep{Winder_19_Mdwarf_multiplicity,Clark_24_binaries}, $\sim 50$\% for FGK stars \citep{RaghavanEtal10}, further increasing with $M_*$ \citep{Offner_23_binaries_review}. The companion frequency distribution peaks at separation of
$a_{\rm b} = 4$ au, 25 au, and 40 au for $0.075 \leq M_*/\msun \leq 0.15$, $0.3 \leq M_*/\msun \leq 0.6$, and $0.75 \leq M_*/\msun \leq 1.25$, respectively \citep[cf.  Table 2 in][]{Offner_23_binaries_review}.

Recent surveys uncovered a strong anti-correlation of close (defined for the purpose of these correlations as $a_{\rm b} < 10$~au) binaries with metal abundance \citep{Moe_Kratter_19_binary_vs_Z,Mazzola_20_binaries_vs_Z}. \cite{Moe_Kratter_19_binary_vs_Z} hence argue that close binaries are made by disc fragmentation since the process is most efficient at low metallicity. This picture is further supported by the binary separations statistics. Secondary objects are expected to form by disc fragmentation at radii $R\gtrsim 100$~au (see \S \ref{sec:initial_fragments} \& \S \ref{sec:fragment_migration}). This separation then shrinks by a factor of a few in general. Simulations of binary systems accreting from their circumstellar discs in the presence of continuous infall of material from larger scales show that the binary separation can shrink rapidly \citep{Artymowicz_91_binaries}, with the secondary often outperforming the primary in terms of mass growth rate \citep{BateB97}. \cite{Tokovinin_Moe_20_binaries} set up a simple but realistic population synthesis study of binary growth through the sequence of multiple uncorrelated growth episodes, and found that it reproduces not only the binary separation and mass ratio statistics, but potentially also the well known brown dwarf desert at small separations \citep[e.g.,][]{Kraus_08_BD_desert}. 
A significant number of eclipsing triple stellar systems is known, and these systems are highly aligned \citep{Rappaport_22_triples,Kostov_24_triples}; thus, as planetary systems, they must have formed in a common disc.



\subsection{FFP formation in the CA scenario}\label{sec:CA_FFP}

\cite{Holman_Wiegert_99} performed long term simulations of planets in binary systems, finding that planets are very vulnerable to ejections. For example, for $q=1$ and mild eccentricity of $e=0.2$, planets orbiting one of the stars are lost unless their $a_{\rm p}\leq 0.2 a$; for circum-binary planets in the same binary, on the other hand, the forbidden zone extends to $a_{\rm p}\approx 3 a$. If $a_{\rm b} = 10$~au, for example, then any planet with semi-major axis from 2 to 30 au is liable to be ejected.

\cite{Coleman-24-FFP-simulations,Coleman-24-FFPs} focus on binaries with separations $a_{\rm b} < 0.5$~au, although they then extrapolate the results to binaries with separation  $a_{\rm b} \leq 3$~au (which are more numerous). Their population synthesis make strong predictions on the mass function of $\mu$FFPs. 
The models are able to match the observed $\mu$FFPs mass function for planets with mass $M_{\rm p}\gtrsim 10\mearth$; however, there is a significant mismatch at lower planet masses. Physically, planets with $M_{\rm p}\lesssim 10\mearth$ are more likely to collide bodily with another planet rather than be ejected \citep[cf. Fig. 3 in][]{Coleman-24-FFPs}. They also grow rapidly in mass, and migrate towards the binary too slowly, and so do not get ejected until they reach the massive super-Earth stage (or not at all). Henceforth, a very strong prediction of their calculations \citep[Fig. 2 in][]{Coleman-24-FFPs} is the strong dearth of FFPs less massive than $\sim 10\mearth$. In fact, the total number of FFPs produced by their calculations in the mass range $0.3 \mearth < M_{\rm p} < 10 \mearth$ is $\sim 1.1$ planet per star, whereas \cite{sumi2023} find $22^{+22}_{-13}$ $\mu$FFPs in the same mass range.

Furthermore, it appears that to account for $\mu$FFPs even above $M_{\rm p}\approx 10\mearth$, close binary systems must be at least an order of magnitude more efficient in hatching solid cores than single stars. $\mu$FFPs appear to roughly rival the population of bound planets, yet the binaries with separation $a_{\rm b}\leq 3$~au represent only $\sim 14$\% of the systems. The planet ejection efficiency of close binaries is $\sim (50-70)$\% \citep{Coleman-24-FFP-simulations}, hence the estimate. The surprising efficiency of solid core formation in binary systems is also unexpected given that the mean metallicity of close binaries is [Fe/H]$\approx -0.2$ \citep{Moe_Kratter_19_binary_vs_Z}. CA predicts that massive solid core formation is unlikely at low metallicities \citep{IdaLin04b,IdaLin08}.

$\mu$FFPs could be bound to stars on orbits with planet semi-major axes $a_{\rm p} \gtrsim 10$ au \citep{Yee-25-FFPs-MFunction}. Such wide separation planets are also very challenging to all flavours of CA: planets more massive than $\sim 10\mearth$ accrete gas and should also migrate in to much smaller separations. Population synthesis studies therefore predict a dearth of such planets beyond 10 au, e.g., see Fig. 2 in \cite{AlibertEtal13}, Fig. 5 in \cite{Bern20-2}, Fig. 3 in \cite{Emsenhuber_23_CA}, or Figs. 7-8 in \cite{Coleman-23-planet-in-CB-systems}. Finally, \cite{ma2016} presented CA population synthesis of single star systems with stellar mass of $0.1\msun$, $0.3\msun$ and $1\msun$, and found that in all cases the number of bound FFPs is too small compared with the numbers of planet ejections in their simulations. This predicted dearth of massive planets at wide separations is also at odds with the abundant bound population of microlensing planets \citep{SuzukiEtal16,SuzukiEtal18}. A similar conclusion was reached by \cite{NduguEtal19}, see their Fig. 6.

\section{Disc fragmentation preliminaries}\label{sec:preliminaries}

\cite{Kuiper51b} proposed that Solar System planets formed by disc fragmentation (GI), followed by evaporation of volatiles. Later work \citep[e.g.,][]{DW75,DeCampliCameron79} concluded that  GI could not have occured inside $\sim 10$~au. However, realisation that planet migration is very important for GI planets \citep[e.g.,][]{VB06,VB10,BaruteauEtal11} led to reincarnation \citep[e.g.,][]{BoleyEtal11a,NayakshinFletcher15} of Kuiper's ideas. Here we propose a framework for formation of both planets and binary stars via disc fragmentation.

\subsection{Initial fragment masses}\label{sec:initial_fragments}


The initial mass of the fragments formed in the disc can be cast as
\begin{equation}
    M_{\rm frag} = A_{\rm frag} M_* \left(\frac{H}{R}\right)^3\;,
    \label{Mfrag_general}
\end{equation}
where $H/R$ is the local disc aspect ratio, $A_{\rm frag}>0$ is a poorly constrained dimensionless factor, with estimates ranging from $A_{\rm frag} \lesssim 1$ \citep[e.g.,][]{BoleyEtal10,Kubli-GI-MHD-23} to $A_{\rm frag}>10$ \citep{KratterEtal10,Xu_25_RHD_disc_fragmentation}. For an irradiation-heated disc orbiting a star of mass $M_*=1\msun$ and luminosity $L_*=1\lsun$, we estimate the effective temperature of the disc via $\sigma T_{\rm eff}^4 = \delta_{\rm irr} L_*/(4 \pi R^2)$ \citep[e.g.,][]{ZhuEtal12a}, and
\begin{equation}
    T_{\rm eff} \approx 20\text{ K } R_{2}^{-1/2}\;,
    \label{Teff}
\end{equation}
where $R_{2} = R/(100$~au) and we set $\delta_{\rm irr} = 0.1$. \cite{Xu_25_RHD_disc_fragmentation} recently presented detailed 3D radiative hydrodynamics simulations, finding that the initial fragment mass, $M_{\rm frag}$, is distributed as a log-normal distribution around a mean. In terms of $A_{\rm frag}$, \cite{Xu_25_RHD_disc_fragmentation} results yield $<A_{\rm frag}> = 40$. Applying eq. \ref{Teff}, their eq. 24 gives the mean
\begin{equation}
    M_{\rm frag} \approx 10\;\mj \left(\frac{A_{\rm frag}}{40}\right)  R_{2}^{3/4}\;,
    \label{Mfrag_Xu}
\end{equation}
with an uncertainty of a factor of $\sim 2$ in both directions. Note that we used here $H = 0.69 c_{\rm s}\Omega_{\rm K}^{-1}$, where $c_{\rm s}$ is the isothermal sound speed, as appropriate for a self-gravitating disc with \citep{Toomre64} parameter $Q\approx 1$. 

\begin{figure}
\includegraphics[width=0.48\textwidth]{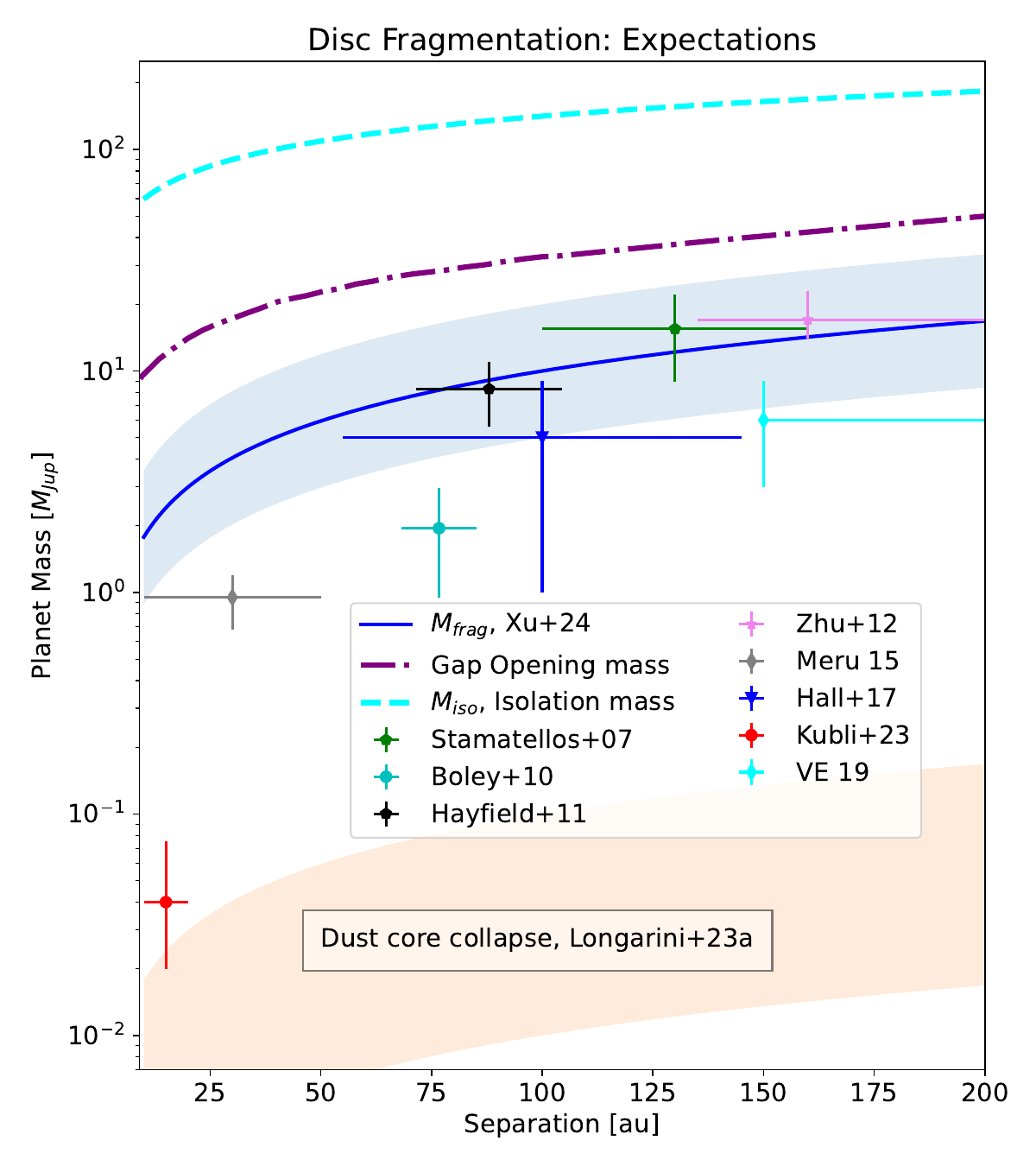}
\caption{The curves show initial fragment mass (\S 
\ref{sec:initial_fragments}), $M_{\rm frag}$, gap opening mass (\S \ref{sec:fragment_migration}), $M_{\rm gap}$, and isolation mass (\S \ref{sec:planets_or_binary}), $M_{\rm iso}$ for a star with mass $M_*=1\msun$. Note that fragment mass is a strongly increasing function of distance.   The points show initial fragment masses from literature, with approximate error bars. The shaded band on the bottom gives the estimated masses of dust cores forming by collapse of the dust component only. }
\label{fig:theory_prelims}
\end{figure}

The blue solid curve in Fig. \ref{fig:theory_prelims} shows $M_{\rm frag}$ from eq. \ref{Mfrag_Xu}, with an associated uncertainty shown as a shaded region around it. We surveyed disc fragmentation literature, and present $M_{\rm frag}$ and fragmentation location for those simulations where these data are either given explicitly or could be deduced from figures. This is shown in Fig. \ref{fig:theory_prelims} by symbols with errorbars \citep[the literature cited in the legend is][]{Xu_25_RHD_disc_fragmentation,Stamatellos_07_BDs,BoleyEtal10,Hayfield_11_GI_disc,HallCEtal17,Kubli-GI-MHD-23,Vorobyov-Elbakyan-19}. Several studies find $M_{\rm frag}$ close to eq. \ref{Mfrag_Xu}. However, others find smaller values.  The most significant outlier in the figure are the results of \cite{Deng_21_GI_Neptunes,Kubli-GI-MHD-23}, where the planets on average weight about one Neptune mass. In contrast to other simulations,  \cite{Deng20-MHD-GI,Deng_21_GI_Neptunes,Kubli-GI-MHD-23} who perform ideal MHD simulations rather than pure hydrodynamics simulations. 

The sand-coloured band on the bottom in Fig. \ref{fig:theory_prelims} shows the range for the estimated masses of dust cores from gravitational collapse of the dust component only  \citep{GibbonsEtal14,Baehr_22_GI_dust,Baehr-23-GI-dust,Longarini_23_solids_collapse_theory}. Here we used the results of \cite{Longarini_23_solids_collapse_sims}, who found that the dust core masses in their simulations of self-gravitating dusty discs are $M_{\rm dust} \sim (10^{-3} - 10^{-2}) \mj$, where $\mj$ is the gas Jeans mass in the disc (for which we use the blue curve from Fig. \ref{fig:theory_prelims}).


\subsection{Fragment migration and gap opening}\label{sec:fragment_migration}

Numerical simulations show that Jovian mass planets migrate inward very rapidly in self-gravitating discs \citep[e.g.,][]{VB06,VB10,BoleyEtal10,MachidaEtal11,ChaNayakshin11a,MichaelEtal11}. \cite{BaruteauEtal11} showed that this migration occurs via a type-I like process \citep[][]{BaruteauEtal14a}. \cite{MalikEtal15} showed that the planet migrates so rapidly that it essentially has no time to open a gap. The migration timescale  is
\begin{equation}
    \frac{t_{\rm mig,I}}{T_{\rm orb}} \sim B_{\rm mig} \frac{h^3}{q_{\rm p}} 
    \Big[ \frac{0.1}{h} \Big]^2\;,
    \label{tmig}
\end{equation}
where $T_{\rm orb}$ is the orbital time, $h=H/R$, $q_{\rm p} = M_{\rm p}/M_*$, and $B_{\rm mig}$ is a dimensionless factor of order unity that depends on the disc properties \citep[see][]{BaruteauEtal11}. For $M_*=1\msun$ and planets of Jupiter mass and above, $t_{\rm mig,I}\sim$ a few thousand years. Planets of mass significantly smaller than Jupiter migrate less rapidly, as the $t_{\rm mig}\propto q_{\rm p}^{-1}$ scaling predicts. At  $M_{\rm p}\gg 1\mj$, \cite{Stamatellos-15,MalikEtal15} found that objects in BD-mass regime are able to open deep gaps in the disc and slow down their migration to the type-II rate, which is one or more orders of magnitude slower. 

Fragments are expected to migrate in type I regime inward until they reach the gap opening mass, $M_{\rm gap}$, which is shown with the purple curve in Fig. \ref{fig:theory_prelims}. 
Relatively low mass objects, i.e., $M_{\rm p} \lesssim 1\mj$ are expected to migrate quite close to the primary star (e.g., the inner few au) before they open a gap. BD-mass objects, on the other hand, are likely to stall at separations of tens of au or more.

\subsection{Mass gain by accretion vs mass loss}\label{sec:accretion_prelim}

Fragments  can gain mass by gas accretion and/or mergers with other fragments \citep{Stamatellos_07_BDs,VB10,KratterEtal10,ZhuEtal12a}, but they can also lose mass due to stellar tides  \citep{BoleyEtal10,Nayakshin10c,ChaNayakshin11a,Vorobyov11a,ForganRice13b}, irradiation from the surrounding hot disc \citep{CameronEtal82,VazanHelled12}. We consider only one massive fragment per disc in this paper, so here we focus on gas accretion.

\cite{ZhuEtal12a} find that many ``planets" in GI discs double in mass on a time scale of a few orbits, and ran away to masses $>0.1\msun$. 
\citep{NayakshinCha13,Stamatellos-15,Nayakshin17a}, however, showed that gas accretion rate onto  planets depends sensitively on gas thermodynamics and planet (clump) mass. \cite{ZhuEtal12a} gas clumps were usually very massive, $M_{\rm p}\gtrsim 20\mj$. In this case essentially all of the gas that enters the Hill sphere of the planet is swiftly accreted; gas accretion rate can be estimated as $\dot M_{\rm p}\sim 4\Sigma R_H^2 \Omega$ \citep[e.g.,][]{Levin07,ZhuEtal12a}, and gas accretion time scale is of order the orbital time \citep[see][]{FletcherEtal19},
\begin{equation}
    \frac{t_{\rm acc}}{T_{\rm orb}} = \frac{M_{\rm p}}{\dot M_{\rm p}} = \frac{3}{8} \frac{R_{\rm H}}{H} Q\;,
    \label{t_acc}
\end{equation}
which is also comparable to the migration time scale. However, gas accretion is strongly suppressed when object mass is too small to bind gas entering its Hill sphere, or when the cooling time is longer than the local dynamical time \citep{NayakshinCha13,Stamatellos-15,Nayakshin17a}. 


On the other hand, at formation, GI planets are as extended as $R_{\rm p} \sim 5$~au in radius \citep{ChaNayakshin11a,GalvagniEtal12,ZhuEtal12a,Vorobyov-Elbakyan-19}. Such planets are more justly called molecular clumps, since they have central temperatures in hundreds of K \citep{Bodenheimer74}, and spend $\sim (10^3-10^5$)~years, depending on fragment mass and opacity, cooling and contracting before they collapse to several orders of magnitude denser post-collapse planets \citep{GraboskeEtal75,BodenheimerEtal80,HelledEtal08,Nayakshin10a,HumphriesEtal19}. These pre-collapse clumps can be tidally disrupted if they migrate closer to the central star \citep[e.g.,][]{Nayakshin10c,ChaNayakshin11a,Vorobyov12,ZhuEtal12a,ForganRice13b}. This occurs when the Hill radius of the planet,
\begin{equation}
    R_{\rm H} = R \left(\frac{M_{\rm p}}{3 M_*}\right)^{1/3} = 7 \text{ au } q_{-3}^{1/3} R_2\;,
\end{equation}
where $q_{-3} = q_{\rm p}/(0.001)$, is $\lesssim$ the planet radius, $R_{\rm p}$.

\subsection{Formation of planets vs binary growth}\label{sec:planets_or_binary}

\begin{figure}
\includegraphics[width=0.48\textwidth]{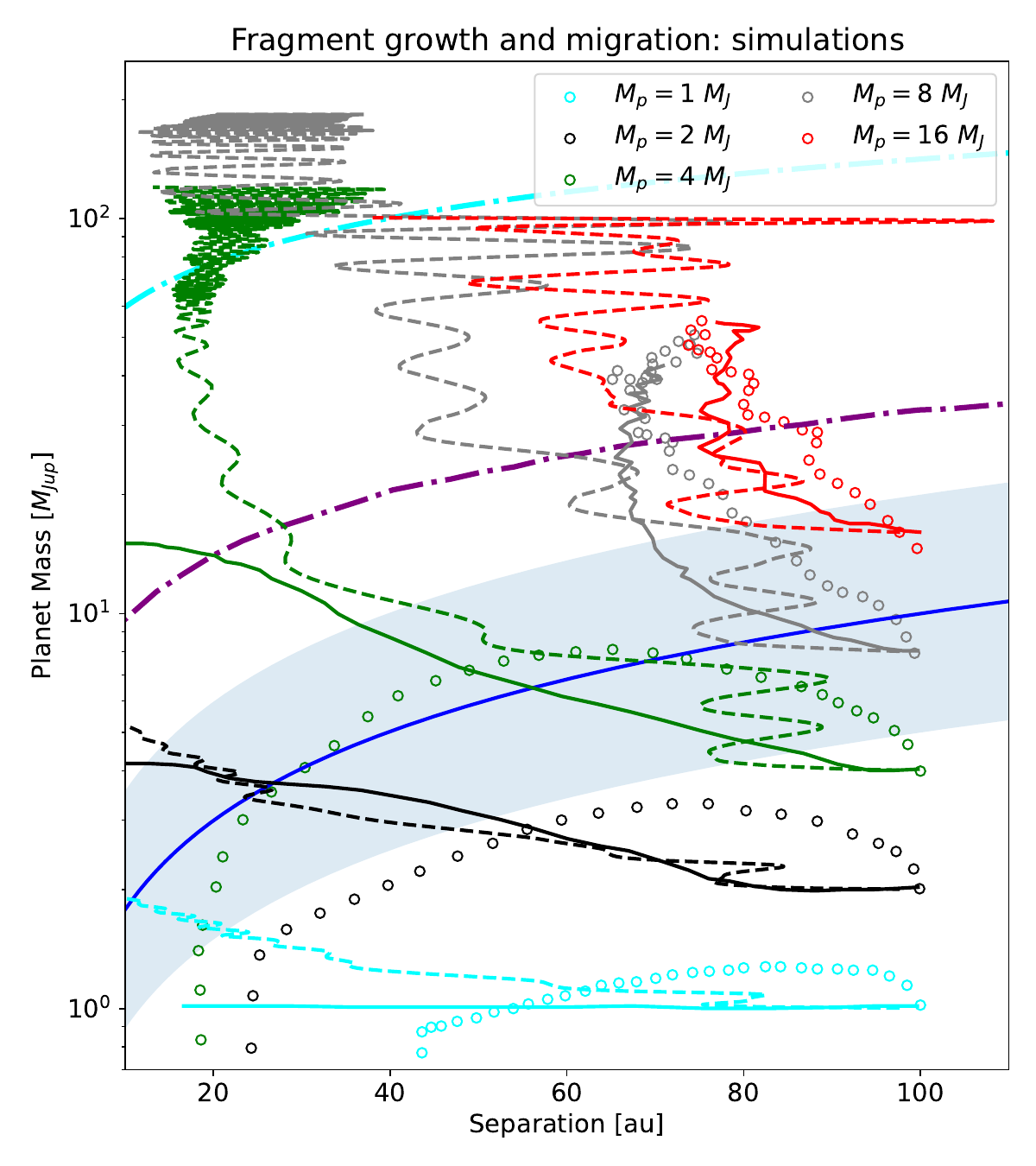}
\caption{Results of three groups of numerical simulations \citep[two from][one from this paper]{Nayakshin17a}, for fragments of four different initial masses (see legend) evolving in massive self-gravitating discs. Note that fragments with mass $M_{\rm frag}\lesssim 4 \mj$ usually remain in the planetary mass regime and migrate inward rapidly. In contrast, higher mass fragments grow in mass quickly, and open a gap. This slows down their migration and allows them to accrete gas in a runaway fashion. Such fragments evolve towards the isolation mass. See \S \ref{sec:preliminaries} for detail.}
\label{fig:simulation_prelims}
\end{figure}


\cite{Nayakshin17a} used 3D SPH simulations to study evolution of a single fragment of a given initial mass, $M_{\rm p}$, injected into a self-gravitating disc at $R=100$~au. Two very different methods of modelling  fragments were used and compared. In the first, the fragments were pre-collapse gas clumps, modelled by physically resolved polytropic gas spheres. In the second, the fragments were assumed to be in the post-collapse configuration, and modelled as a sink particles \citep{Bate95}. Fig. \ref{fig:simulation_prelims} presents the resulting tracks that the planets made in the mass-separation parameter space with dots (pre-collapse planets) and solid curves (sink particle planets), respectively. Five values of $M_{\rm frag}$ are shown in Fig. \ref{fig:simulation_prelims}. \cite{Nayakshin17a} used the same $M_*=1\msun$, and the same irradiation temperature profile as we use here (eq. \ref{Teff}), so the three curves from Fig. \ref{fig:theory_prelims} -- $M_{\rm frag}$, $M_{\rm gap}$ and $M_{\rm iso}$ -- are the same and shown in Fig. \ref{fig:simulation_prelims} for reference again. 


In Fig. \ref{fig:simulation_prelims}, planets with initial $M_{\rm p} = (1-4)\mj$ accrete gas slower than they migrate, and therefore they stay below $M_{\rm gap}$, and migrate inward rapidly. What happens to these planets depends on whether they are in the pre-collapse or the post-collapse configuration. In the former case, planets are disrupted. Although dust was not included in these simulations, it is expected that dust sedimentation to the planet centre will make a solid core there \citep[e.g.,][]{Kuiper51b,Boss97,Boss98,BoleyEtal10,ChaNayakshin11a}. This may result in release of dust cores of mass up to $\sim 10-20\mearth$ \citep{Nayakshin16a} back into the disc at $20-40$~au. Where exactly this disruption occurs, and whether the cores are able to hold on to some of the clump gas, is very model and parameter dependent  \citep[see, e.g.,][]{Nayakshin15c,Nayakshin16a}, but potentially this scenario may result in all sorts of sub-Jovian planets formed \del{very rapidly} at a wide range of radii \citep{NayakshinFletcher15} very quickly (i.e., faster than 0.1 Myr). If the fragments are in the post-collapse configuration, then they may migrate inside $R\leq 10$~au and remain gas giants. 


In Fig. \ref{fig:simulation_prelims}, more massive fragments are able to accrete gas efficiently. These fragments grow into massive BDs and would have, probably, grown further if the simulations were continued for longer, thus forming secondary stars\footnote{Simulations of binary formation that include  external mass deposition show that, usually, the mass ratio of the binary system increases with time \citep{BateB97,Young_Clarke_15_binaries}, indicating that it is the secondary that grows in mass the most.}. 


We argue that these results are robust, at least qualitatively. \cite{Nayakshin17a} used the 3D SPH code \texttt{GADGET} \citep{Springel05} and an approximate radiative cooling scheme\footnote{Curves shown in Fig. \ref{fig:simulation_prelims} correspond to the nominal dust opacity factor, $f_{\rm op} =1$, in the quoted paper.}. In \S \ref{sec:FARGO} we present 2D fixed grid hydrodynamics code \texttt{FARGO-ADSG} simulations. We use a one-zone approximation to the radiative cooling, which makes our \texttt{FARGO-ADSG} setup physically similar but numerically quite distinct that of \cite{Nayakshin17a}. We include external mass deposition in our disc in a manner similar to \cite{ZhuEtal12a}, which makes our disc initial surface density different from the SPH calculations in Fig. \ref{fig:simulation_prelims}. Finally, our treatment of gas accretion onto the planet is sink-particle like \citep[a modified version of ][]{Kley_99_planet_accretion}, but again its numerical implementation is very different from that of \cite{Nayakshin17a}. For the purposes of a comparison with \cite{Nayakshin17a}, we injected sink-particle like objects into the disc at $R=100$~au and they were let go immediately. The dotted planet evolution tracks in Fig. \ref{fig:simulation_prelims} show the results for the same initial planet masses. We observe that there is a qualitative agreement in the results of \texttt{FARGO-ADSG} simulations with that of \texttt{GADGET}. For the dashed curve tracks, the $M_{\rm p} = (1-2)\mj$ planets migrate in rapidly, although gaining somewhat more mass than the sink-particle planets did in \cite{Nayakshin17a}. The $M_{\rm p}=4\mj$  \texttt{FARGO-ADSG} planet opened a deep gap in the disc and then settled on the runaway gas accretion track towards the isolation mass, whereas \cite{Nayakshin17a} planets did not open the gap.
The cases of $M_{\rm p} = (8-16)\mj$ planets are also  qualitatively similar between the three groups of simulations; they runaway in mass by gas accretion and then stall on relatively wide orbits. One also notes that \texttt{FARGO-ADSG} planets become low mass secondary stars.

\subsection{Other scenarios for formation of small planets}\label{sec:MHD_or_dust_prelim}

In \S \ref{sec:accretion_prelim} \& \S \ref{sec:planets_or_binary} we explored ideas following the ``Tidal Downsizing" model of \cite{Nayakshin10c,Nayakshin_Review}. There are two other ways to form low mass planets via disc fragmentation. 

\cite{RiceEtal04} performed 3D SPH simulations of gas-particle discs  and found that spiral density arms can be effective particle traps for dust grains with Stokes number $\text{St}\sim 1$. The local density enhancements were sufficiently large for the dust density to exceed that of the gas and potentially lead to formation of particle clumps with masses as high as $10\mearth$ or more. \cite{GibbonsEtal12,Shi_16_GI_dust_focusing,BoothClarke16} investigated dust velocity dispersions in 2D simulations of self-gravitating discs, finding that grain growth up to $\text{St}\sim 0.1-1$ is indeed possible within spiral density arms due to dust velocity being highly correlated. \cite{GibbonsEtal14} found localised dust collapse to particles with mass up to $0.01\mearth$ in their shearing box simulations. More recently, \cite{Longarini_23_solids_collapse_theory} derived conditions for dust particle collapse within spiral density arms in self-gravitating discs analytically, and \cite{Longarini_23_solids_collapse_sims} used 3D SPH simulations to study the problem numerically, confirming that solid cores with masses as high as $10\mearth$ can form \citep[see also][]{Baehr_22_GI_dust,Baehr-23-GI-dust}. \cite{Rowther_24_GI_dust} found collapsed dust concentrations masses in the range $\sim 0.15- 6\mearth$. In these works, the gaseous disc is self-gravitating but is non-fragmenting.

Inclusion of magnetic fields in self-gravitating disc simulations leads to the development of a novel turbulent dynamo driven by vertical flow circulation across spiral arms which supersedes the MRI \citep{Riols_17_GI_MHD,Riols_18_GI_dynamo,Riols_19_GI_dynamo,Deng20-MHD-GI,Bethune_22_GI_dynamo}. \cite{Deng_21_GI_Neptunes,Kubli-GI-MHD-23} perform 3D MHD simulations of self-gravitating discs and find the formation of self-bound {\em gaseous} gas clumps with mass as low as Super Earths to Neptune-sized  planets, and radii of only tens of au. The latter simulations could not follow the pre-collapse phase of these low mass clumps, but preliminary work suggest that they would range from as little as $10^4$ yr to a million years (Beffa, Kubli et al., in preparation). 

\section{Our FFP model and toy calculations}\label{sec:motivation_prelim}

\subsection{The model}\label{sec:our_model}

Our key assumptions are:

\begin{enumerate}
    \item[(A)] Planets are born at relatively small separations $R\lesssim$ tens to $\sim 100$~au, where $M_{\rm frag}$ is lower (cf. \S \ref{sec:initial_fragments}). Fragments of lower mass cool less rapidly, and present more hospitable environment for formation of solid cores \citep{HS08,Nayakshin10a,Humphries-21-GI-ALMA}. 
    \item[(B)] The seed of the secondary star is born by disc fragmentation at larger separations, $R \gtrsim 100$~au, or at least at a later time. This is logical since $M_{\rm frag}$ increases with $R$ (Fig. \ref{fig:theory_prelims} and \S \ref{sec:initial_fragments}); also, gas radiative cooling is more rapid at large $R$, allowing more rapid gas accretion onto the seed (\S \ref{sec:accretion_prelim}). Disc fragmentation at larger radii may be occur later in time because protoplanetary discs grow inside-out \citep{VB06,VB10,VB15}. In this picture, by the time the secondary object is born, the planets are already present in the disc.
    \item[(C)] Gas accretion onto planets is negligible whereas the secondary star grows in mass extremely rapidly. This follows from the discussion in \S \ref{sec:accretion_prelim}.
\end{enumerate}

Unlike the CA scenario, the secondary in the disc fragmentation scenario does {\em not} stifle formation of smaller planets because in our model, the secondary forms later. At the time of smaller planet/solids formation, there is no difference in the disc structure between single and multiple stellar systems. The distinction is that in binaries the discs are larger and/or more massive. Secondary objects were shown by \cite{Meru15,Cadman_22_triggered_fragm_binary} to trigger disc fragmentation, helping not hindering planet formation.

 After the secondary star formation, the binary system shrinks \citep{Tokovinin_Moe_20_binaries}, resulting in a rapid ``pincer movement". Low mass planets often find themselves in the instability zones deliniated by \cite{Holman_Wiegert_99}, and are henceforth ejected as FFPs.




\subsection{Critical mass of the migrating secondary}\label{sec:pincer}

We will see below that planet ejection via close interactions with the secondary is related to the  ``gravitational assist" effect  (see 
\href{https://en.wikipedia.org/wiki/Gravity_assist}{https://en.wikipedia.org/wiki/Gravity\_assist}) from Solar System exploration studies. Here we present order of magnitude analytical picture of how these ejections work.

Compared with the two stars and the disc, the low mass planet we consider is essentially a test particle. It is convenient to study the interaction in the moving frame of the secondary, which we assume much less massive than the star for simplicity. Neglecting the role of gas during a close planet-secondary interaction, in the secondary's frame, the planet enters and leaves its vicinity with exactly the same relative velocity $|\Delta {\bf v}|$, except for a change in direction of $\Delta {\bf v}$. However, planet's velocity in the frame of the star, ${\bf v_p}$, changes. If the change reduces $|{\bf v_p}|$, the planet gets on a more bound orbit; in the opposite case an ejection is possible. 

The most extreme example of the effect is provided by the planet and the secondary both moving on circular orbits but in opposite directions, one prograde and the other retrograde. The planet velocity in the secondary's frame is hence $2 v_{\rm c}$ before and after the collision. Let us assume that the planet loops around the secondary and comes back with an exactly opposite velocity, which is again $2 v_{\rm c}$ in magnitude. In the frame of the star the planet velocity is now $v_{\rm c}$ (secondary's velocity) plus $2 v_{\rm c}$, hence $3 v_{\rm c}$. This planet is then ejected because the local escape velocity is $\sqrt{2} v_{\rm c}$. This example is not realistic at all for the problem at hand, but it shows that planet scattering on a significant angle in the frame of the secondary can lead to an ejection.

In the Solar System studies, spacecraft approaches a planet on a fixed orbit. Here the opposite is true. The low mass planet migrates slowly, so is approximately on a fixed orbit, whereas the secondary moves in on it rapidly from the outer disc. We now show that if the secondary is of a sufficiently low mass, then it is quite likely for it to zoom past the low mass planet without ejecting it.

We assume that both the planet and the secondary are on approximately circular orbits with radial velocity given by $v_{\rm mig} = -R/t_{\rm mig}$, where $R$ is the radius of the orbit. For the planet, we neglect radial migration, whereas for the secondary we do not.
For an ejection event, we require the planet's velocity in the secondary's frame to change in direction by a large angle, which we set to be $\geq \pi/2$. Classical mechanics' hyperbolic orbit result then requires that the interaction impact parameter is $\Delta R_c = GM_s/\Delta v^2$, where $\Delta v$ is the relative velocity between the planet and the secondary.

Such a close encounter is only possible when the secondary is inside a ring with width $2 \Delta R_c$ centred around the orbit of the planet. The time it takes the secondary to cross this region due to migration is
\begin{equation}
    t_{\rm cross} = \dfrac{2 \Delta R_c}{v_{\rm mig}} = 2 t_{\rm mig} \dfrac{\Delta R_c}{R}.
\end{equation}
Further,  the secondary and the planet must get close enough in the azimuthal direction to experience strong scattering. In the frame of the planet, the relative azimuthal velocity is $\Delta v_y = -\dfrac{3}{2} \Omega x$, where $\Omega$ is the angular frequency of the orbit, and $x$ is the radial distance between the planet and the secondary. Note that $\Delta v_y$ changes signs as the secondary crosses $x = 0$, which means that the secondary orbits slower than the planet outside, and faster inside.

For there to be a close approach, the secondary must be able to cover an azimuthal distance of no less than $2\pi R$ between $x = \Delta R_c$ and $x = 0$, since the azimuthal catch-up can only happen in one direction. This distance is given by $\Delta y = \int_{x=\Delta R_c}^{x=0} \Delta v_y \,dt$. Substituting in $\dfrac{dx}{dt} = - v_{\rm mig}$, we integrate and find:
\begin{equation}
    |\Delta y| = \dfrac{3}{8} \Omega t_{\rm cross} \Delta R_c
    = \dfrac{3}{4} \Omega t_{\rm mig} \dfrac{\Delta R_c^2}{R},
\end{equation}
A close approach occurs when $|\Delta y| > 2\pi R$. We parametrise the migration time in units of the local orbital time, $2\pi/\Omega$, as $t_{\rm mig} = N_{\rm mig} (2\pi/\Omega)$, where $N_{\rm mig}$ is a few to few tens. Finally, we get:
\begin{equation}
    \dfrac{3}{4} N_{\rm mig} \Bigl( \dfrac{GM_s}{\Delta v^2} \Bigr)^2 > R^2.
\end{equation}

Simulations show that both the secondary and the planet can develop significant eccentricities due to interactions with spiral density arms (e.g., see Figs. \ref{fig:pla-2D} \& \ref{fig:sph_result}). Hence we write $\Delta v = \xi v_{\rm kep}$, where $\xi < 1$ is a dimensionless parameter, and $v_{\rm kep}$ is the Keplerian velocity of the planet. In general, $\xi \sim 0.2$ appears reasonable based on our simulations below. This yields
\begin{equation}
    q_{\rm crit} = \dfrac{M_s}{M_*} > \sqrt{\dfrac{4}{3N_{\rm mig}}} \,\xi^2 = 0.02\; \left(\frac{5}{N_{\rm mig}}\right)^{1/2}\; \left(\frac{\xi}{0.2}\right)^{2}\;.
\end{equation}
We expect that objects with $q\ll q_{\rm crit}$ will rarely eject low mass planets, whereas the reverse is true in the opposite limit.

Finally, this can be taken further in the laminar disc migration limit. If both orbits stay circular, and the migration is smooth instead of stochastic, then $\Delta v = v_{\rm mig}$, therefore $\xi_{\rm lam} = N_{\rm mig}^{-1}$. Substituting in gives $q_{\rm crit} \approx N_{\rm mig}^{-5/2}$ in this limit.

\subsection{Simple N-body experiments}\label{sec:toy}

\begin{figure*}
\includegraphics[width=0.48\textwidth]{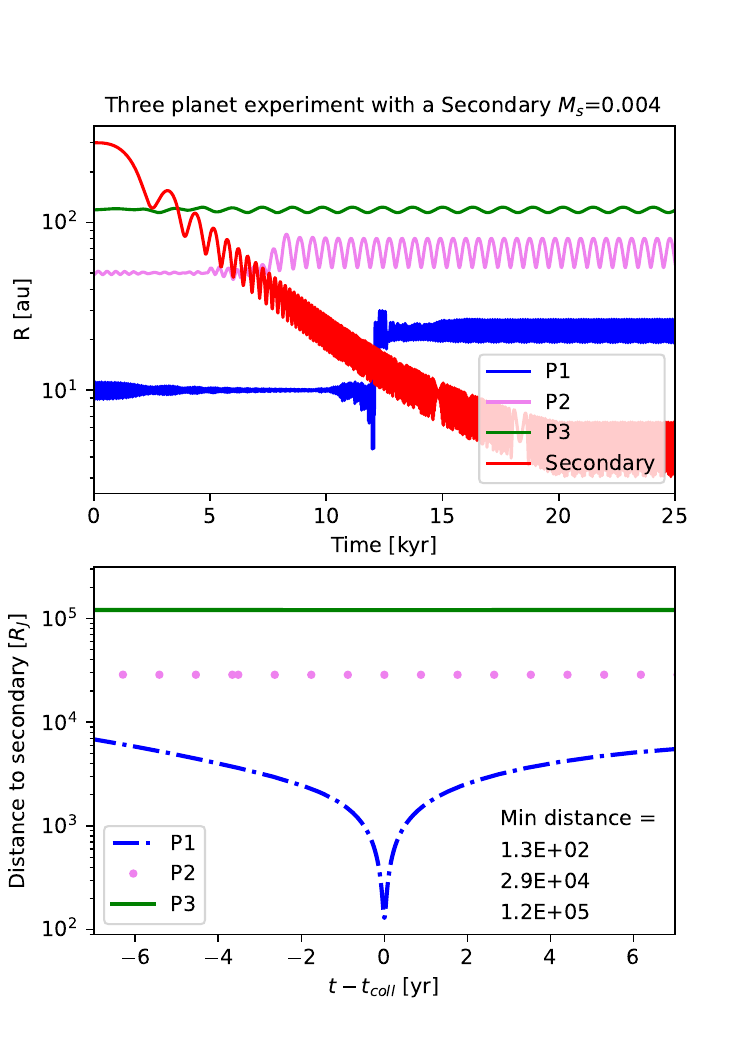}
\includegraphics[width=0.48\textwidth]{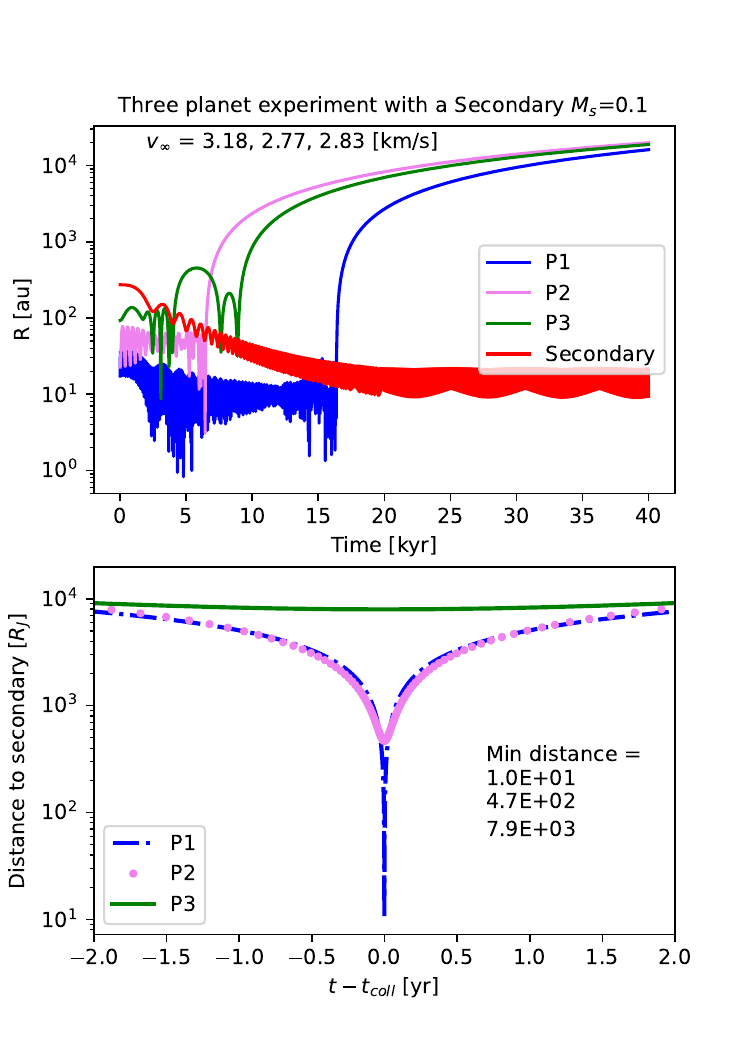}
\caption{The toy secondary object migration experiments described in \S \ref{sec:toy}. Three low mass non-migrating planets are strongly affected by interactions with the secondary. Top panels: Separation of the secondary and the planets to the central star versus time. Bottom panels: Separation between the planets and the secondary object vs time (which is in years rather than kyrs). Left: A planetary mass ($M_{\rm s}\approx 4\mj$) secondary pumps eccentricity but does not eject the planets. Right: A BD-mass ($M_{\rm s}\approx 50\mj$) secondary imparts larger velocity kicks to the planets, all of which are ejected.}
\label{fig:toy}
\end{figure*}

Here we present N-body calculations in which the disc presence is felt only by the migrating secondary interacting with three non-migrating low mass solid bodies. Except for the (prescribed) migration torques on the secondary, the disc gravitational potential is ignored. We use \texttt{REBOUND} \citep{Rein_12_Rebound} to solve the N-body problem with the integrator `ias15' used \citep{Rein_15_Rebound_ias}, as recommended, to deal with close interactions. In the test, a massive secondary migrates in, starting from a circular orbit with an initial ``binary separation" $a=300$~au. The migration is simulated with an exponential frictional drag law:
\begin{equation}
    \frac{d{\bf v}}{dt} = - \frac{{\bf v}}{\tau_a}\;,
    \label{dvdt}
\end{equation}
where $\tau_a = \tau_0 e^{a/R_{\rm stall}}$, and $\tau_0=7$~kyr. Such a rapid migration is observed in simulations of massive planet migration in self-gravitating discs \citep[e.g.,][]{BaruteauEtal11,FletcherEtal19}. In equation above,  $R_{\rm stall}$ is the ``stalling radius" introduced to mimic the results of hydrodynamical simulations of planet/BD migration, which show that massive objects eventually open gaps and switch to the much slower type-II like migration \citep[e.g.,][]{MalikEtal15}. Also, \cite{Rowther_20_stalled_GI_migration} show that planet migration often stalls in the inner disc at $R\sim 10$~au where the disc radiative cooling is inefficient and the disc becomes non self-gravitating. We set $\tau_0=\infty$ at time $t > 20,000$~years to model protoplanetary disc dispersal.

In addition to the secondary object, there are three low-mass planets, $M_{\rm p} = 0.1\mearth, 0.3\mearth$ and $1\mearth$ on initially circular orbits at  $a_{\rm p}=10$, $50$ and $120$~au, respectively. The central star mass is $1\msun$. The masses of the planets are very low to affect the dynamics of the system on the short time scales considered, so they effectively are test particles.

Two representative cases are considered, $M_{\rm s} = 0.004\msun$ and $0.1\msun$. As massive objects open gaps and stall sooner, we set $R_{\rm stall} = 5$~au and $25$~au, for the two respective cases. Fig. \ref{fig:toy} shows the resulting object-star separation evolution (top panels) and planet-secondary distance (bottom panels, given in Jupiter radii, $R_{\rm J}$). In the bottom plots, the time axis is shifted for each planet, so that $t=0$ corresponds to the time of the closest approach between the respective planet and the secondary. Also note that the $t-t_{\rm coll}$ time axis is in years, not kyrs. To monitor that carefully, the ``collision resolve" feature of \texttt{REBOUND} is employed to record the planet-secondary distance data separately from the courser model output. 

We observe that the $4\mj$ secondary object (left panels) perturbs low mass planets' orbits, exciting eccentricities, as expected \citep[e.g.,][]{juric2008}. These orbital perturbations are weaker for planets in wider orbits because the planet-secondary distance of the closest approach (bottom left panel in Fig. \ref{fig:toy}) correlates positively with the orbital radius of the planet. Since $t_{\rm mig}$ is constant in our toy model, the secondary spends more time in units of orbital time at smaller radii, and hence has a greater chance for a closer encounter with a planet there.  Note that the massive planet ended up interior to less massive ones even though it started at larger $R$.

On the other hand, stellar-mass secondary ejects all three planets (the right panels in Fig. \ref{fig:toy}). Planet velocities at infinity, $v_\infty$, are listed in the top right panel of Fig. \ref{fig:toy}, and are all around 3~km/sec in this case, a fraction of the escape velocity from tens of au where the interaction took place. What is interesting is that $v_\infty$ is not correlated with the distance of the closest approach to the secondary (the bottom right panel of Fig. \ref{fig:toy}). This shows that while quite close interactions between the secondary and the planet are possible (planet P1 would merge with the secondary if the secondary's radius is larger than $10 R_{\rm J} \approx 1 \rsun$; this is not unreasonable for a rapidly accreting protostar), encounters do not have to be particularly close to eject the planets. This is because the secondary is massive. The relevant velocity scale for the interaction, 
\begin{equation}
    \Delta v\sim \left[\frac{GM_{\rm s}}{\Delta r}\right]^{1/2} = 3 \text{ km/s} \left[\frac{M_{\rm s}}{0.05 \msun}  \frac{5 \text{ au}}{\Delta r}\right]^{1/2}\;,
    \label{delta_v}
\end{equation}
is comparable to the velocity kick required to unbind the planets, provided that $\Delta {\bf v}$ is in the right direction.

To somewhat quantify the statistical significance of these experiments, we repeated these calculations 10 times, this time for three values of $q = 0.004$, 0.02 and 0.1. The initial azimuthal positions, $\phi_i$, of the planets and the secondary sampled the uniform random distribution from $\phi_i=0$ to $\phi_i=2\pi$. We assumed a perfect merger for planets approach the secondary closer than $\Delta R = 1\rj$, but none of the planets experienced that. The ejection fractions obtained were $F_{\rm ej} = 0.033, 0.27, 0.47$ for $q = 0.004$, 0.02 and 0.1, respectively. The fraction of bound planets is $1-F_{\rm ej}$. There is also a significant trend in the experiments that the surviving planets are on wider, less bound, orbits for larger $q$. In a realistic setting, stellar flybys or tides may unbind the most loosely bound planets this implies that $F_{\rm ej}$ by massive secondaries, $q> 0.1$, can be very high, i.e., well exceeding 0.5.

\section{A 3D SPH calculation example}\label{sec:phantom}

\begin{figure}
    \centering
    \includegraphics[width=0.45\textwidth]{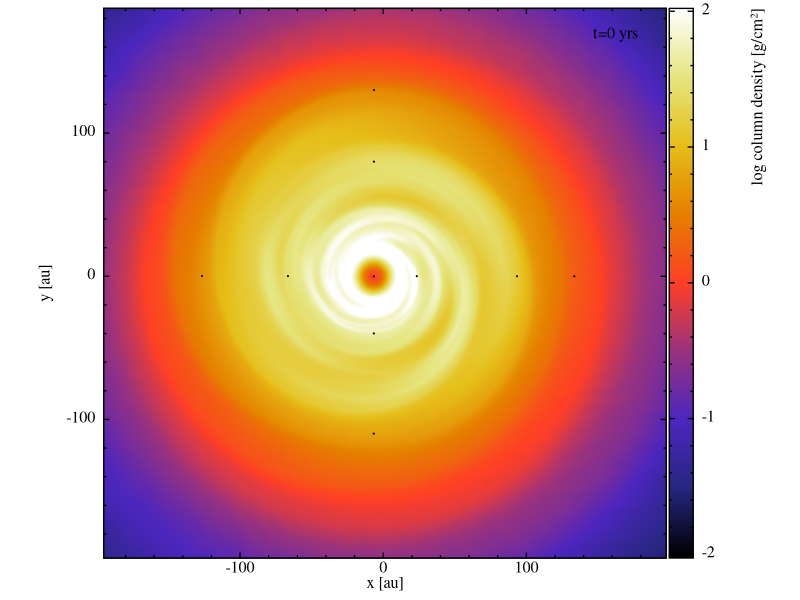}
    \caption{The initial configuration of planets in the \textsc{phantom} simulation. The secondary has a mass of 25 $\mj$ while the 8 planets have masses of 3 $\times$ 10$^{-3}$ $\mj$.}
    \label{fig:intial3D}
\end{figure}

\begin{figure}
    \centering
    \includegraphics[width=0.45\textwidth]{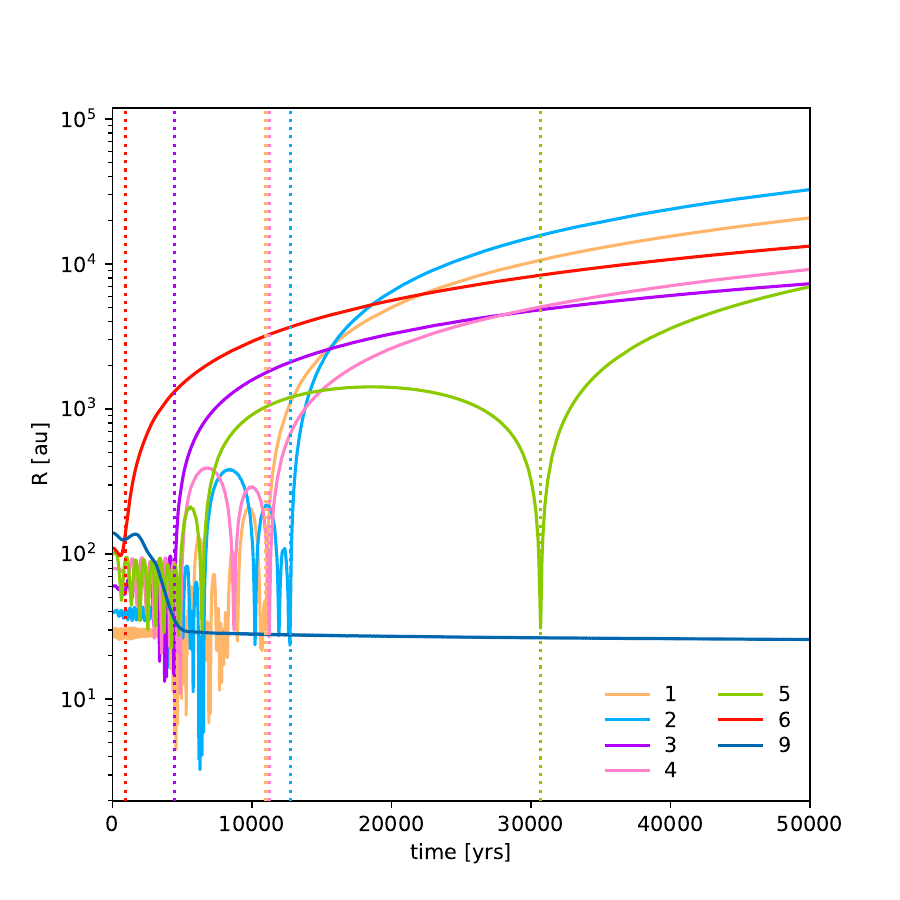}
    \includegraphics[width=0.45\textwidth]{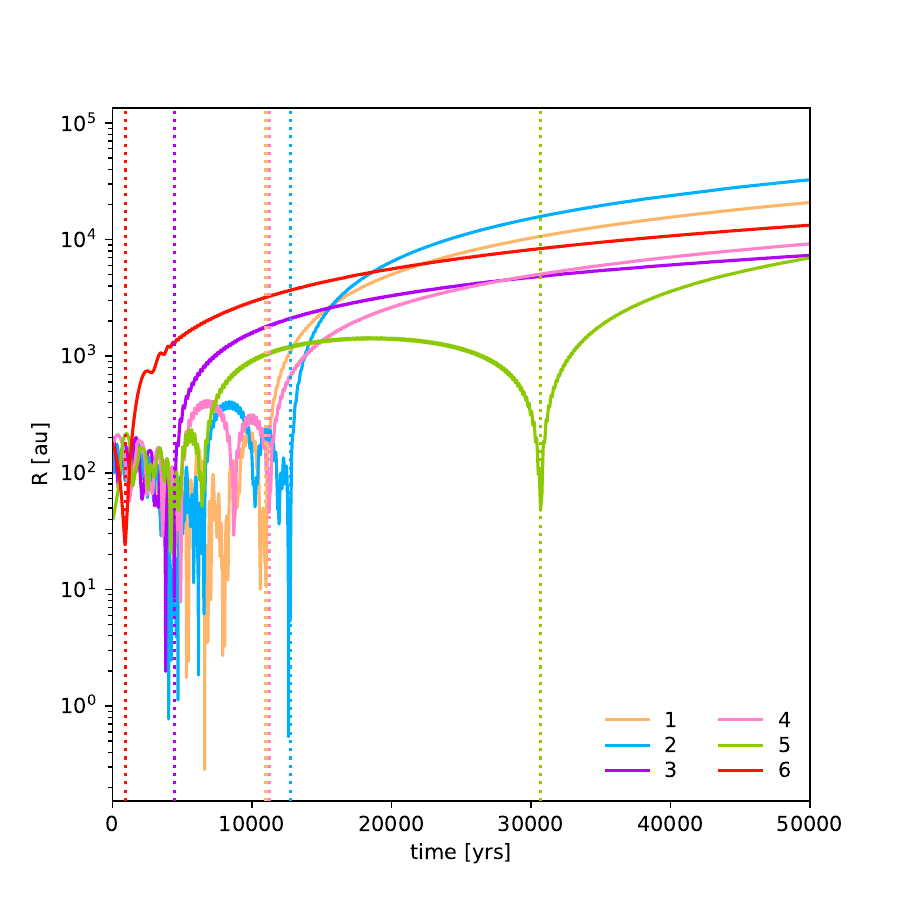}
    \caption{{\em Top panel}: the star-planet distances for all of the ejected planets in the \texttt{Phantom} simulation, with planet ID given in the legend. The secondary (ID equal to 9) starts with $M_{\rm s}=25 \mj$ and grows in mass to $\approx 50\mj$ by the end of the simulation. Note how rapidly it migrates from its initial $R=140$~au orbit to $R\approx 30$~au, where it stops by opening a gap, and consuming most of the local gas there. {\em Bottom panel}: Planet-secondary distances versus time. Note that for some of the planets, the distances of the closest approach are rather large.}
    \label{fig:sph_result}
\end{figure}

\cite{Calovic_25_FFP-1} presented 3D smoothed particle hydrodynamics (SPH) code \textsc{phantom} \citep{Price18-Phantom}  study of $M{\rm p}=(1-3)\mj$ mass planet ejections from a self-gravitating disc with an embedded massive secondary object. They find the process very efficient, with up to $2/3$ of planets ejected as long as the secondary is more massive than $M_{\rm s}\sim 25\mj$. Here we use the same numerical framework except for considering lower mass planets, and we therefore overview the simulation setup only briefly. We start with a massive, $M_{\rm D} =$ 0.32 $M_*$, self-gravitating disc  orbiting a $M_* = 0.5\, \msun$ star. The disc is represented by $n_{\rm p} = 10^6$ SPH particles, and is relaxed before the planets and the secondary object are injected into it on initially circular orbits. As is frequently done \citep{Rice05,BaruteauEtal11,MeruBate12,Longarini_23_solids_collapse_sims}, we use the $\beta$-cooling parameter to describe disc cooling, in particular $\beta = 30$ here, so that the disc is in a gravito-turbulent self-regulated state \citep{Gammie01} but does not fragment during the simulations.

Once the disc settles into the self-regulated state, we insert 8 planets, all with masses of $1\mearth$ at distances of 30 -- 130 au from the star \citep[cf.][]{Calovic_25_FFP-1}, and with initial azimuthal angle given by $\phi_i = (i-1) (\pi/2)$, where $i$ is the planet number. As with \texttt{FARGO-ADSG} simulations presented in \S \ref{sec:FARGO}, using this many planets is motivated by economy of numerical resources (one of our \texttt{PHANTOM} run takes up to 3 weeks on 24 CPUs), and by the fact that planet masses are tiny compared with that of the disc and the secondary, so that they are effectively in the test particle approach. In \cite{Calovic_25_FFP-1} we checked that main results remained the same if only 2-4 planets per disc were used. We also embed a secondary object with a mass of $M_{\rm 2} = $ 25 $\mj$ in the outer disc on orbit at 140 au. The simulation is then run for 95,000 years. This initial configuration can be seen in figure \ref{fig:intial3D}, and is exactly the same as in \cite{Calovic_25_FFP-1}. The star, the secondary and the planets are all modelled as sink particles but with various accretion radii. Therefore, the planets in the simulations presented here do not accrete gas at all, whereas the star and the secondary's mass increase from $M_*=25\mj$ to $M_{\rm s}\approx 50\mj$ by the end of the simulation.

As expected, the secondary migrates inward very rapidly, so that the star-secondary separation shrinks from 140 to 20 au in just 4,000 years \citep[cf. Fig. 3 in][]{Calovic_25_FFP-1}. The top panel of Fig. \ref{fig:sph_result}  shows star-planet distance versus time for all of the objects that were ejected (planets with numbers 1 -- 6 in the legend) plus the secondary (object 9, dark blue line). We do not include planets that were not ejected (7-8) for figure clarity. The ejection times are denoted with the dotted vertical lines. The ejection fraction for this simulation is 75\%. Most of the ejections occur early on, when the violent orbital reconfiguration occurs due to the rapid migration of the secondary. However, planet 5 (green line in figure \ref{fig:sph_result}) is not ejected immediately. It undergoes several close encounters with the secondary, which pump its eccentricity, almost unbinding it. Planet 5 then comes back into the inner disc and is ejected at $\sim 30,000$ years. We observe that not only planet 5 but many of the planets experience more than one close interaction with the secondary before their ejection.

The bottom panel of Fig. \ref{fig:sph_result}  shows planet-secondary distance versus time. In general, the closest interaction of a planet with the secondary is the one that ejects it; however, this is not always true, as is the case with planet 1 in the figure. As remarked in \S \ref{sec:pincer}, depending on the exact nature of the interaction, some planet-secondary scatterings take energy away from the planet, and it may take more than one close interaction to eject the planet.

\section{2D simulations}\label{sec:FARGO}

\subsection{General approach}\label{sec:fargo_general}

Here we present 2D hydrodynamical simulations of self-gravitating disc evolution  with embedded low mass planets and a growing secondary object. We use the 2D fixed grid hydrodynamical code \texttt{FARGO-ADSG}, an extension of the original \texttt{FARGO} code \citep{Masset00}. The code implements both self-gravity and radiative cooling \citep{Baruteau_Masset_08_radiative_disc,Baruteau_Masset_08_typeI,Marzari_12_circumbinary_discs} with the \cite{Bell94} opacity, and the von Neumann-Richtmyer artificial viscous heating.

\cite{ZhuEtal12a} used \texttt{FARGO-ADSG}  to simulate fragmentation of protoplanetary discs in the presence of external mass deposition, at a rate $\dot M_{\rm dep}$, into an annulus centred on radius $R_{\rm dep}$. Simulating discs for a range of $\dot M_{\rm dep}$ and $R_{\rm dep}$,  \cite{ZhuEtal12a} found the disc fragmentation boundary , the critical mass deposition rate, $\dot M_{\rm frag}(R_{\rm dep})$, and then studied evolution of the fragments if/when the disc fragmented. The fragments migrated in, generally, grew in mass by gas accretion and/or lost mass via tides and shocks from spiral arm passages, and were destroyed. Some of the fragments grew as massive as $0.3 \msun$, indicating that binary stars can indeed be formed from first principles in the setup we explore here. 

We use an Equation of State (EOS) with a fixed adiabatic index $\gamma=1.4$ for the disc. As \cite{ZhuEtal12a} noted, due to this simplification, fragments have physical sizes of a few au even when they become sufficiently hot in their centre to collapse to sizes $\sim 100$~times smaller via H$_2$ dissociative collapse of First Cores/gas giant planets \citep{Larson69,Bodenheimer74,BodenheimerEtal80}. As we are interested in low mass solid cores, we model the planets and the secondary as sink/N-body particles. \texttt{FARGO-ADSG} uses the 5th order Runge-Kutta scheme to treat N-body interactions. This scheme uses the same time-step as the code hydrodynamical step. To increase the accuracy of the N-body integrator, we use sub-cycles for N-body interactions, making 32 sub-steps of time duration $1/32$ of the hydro timestep of the code. We also soften N-body interactions with gravitational softening parameter equal to $\varepsilon_{\rm s} = 10^{-4} R_{\rm s}$ where $R_{\rm s}$ is the star-secondary separation. As an example, for $R_{\rm s}=100$~au, this corresponds to $\varepsilon_{\rm s}\approx 20\rj$. We expect our integrator to be adequate for our simulations, since in \S \ref{sec:toy} we found that interactions within $\sim$ a few au are often sufficient for planet ejections (cf. Fig. \ref{fig:toy}); also, \texttt{REBOUND} experiments show that interactions with the minimum distance smaller than $20 \rj$ are very rare.

\subsection{The fiducial run}\label{sec:fargo_setup}

We explained in \S \ref{sec:preliminaries} that the microphysics of disc fragmentation is still not fully understood. Therefore, here we do not attempt to simulate the exact mode by which the disc fragments. As with \texttt{PHANTOM} simulations (\S \ref{sec:phantom}), we create an initial condition in which the disc is close to the fragmentation boundary, inject planets and a (seed) secondary object, and explore the dynamical evolution of the system.

We use open boundary conditions at both the inner and the outer grid boundaries, with a logarithmic grid between $R_{\rm in} = 2$ and $R_{\rm out} = 1000$~au. The number of cells is set to 500 in both radial and azimuthal directions, which results in approximately square grid cells with linear size $\Delta R/R = 0.0124$, which is sufficient to study disc fragmentation according to \cite{ZhuEtal12a}. The calculations are performed in a non-inertial frame centred on the primary star \citep{Baruteau_Masset_08_typeI,Baruteau_Masset_08_radiative_disc}, and co-rotating with the secondary. The disc self-gravity and planet-disc interactions are softened with softening parameters equal to $0.6 H$ and $0.5 H$, respectively, where $H=c_s/\Omega_K$ is the disc local scale-height. 

To create a disc close to fragmentation boundary, we first run a disc relaxation with no embedded objects. At $t=0$, we start with a massive gas disc orbiting a star of mass $M_*=1\msun$ with the initial surface density, $\Sigma(R)$, given by 
\begin{equation}
    \Sigma(R) = \Sigma_0 \left(\frac{R}{R_{0}}\right)^{-1.5} \exp\left[-\frac{R}{R_0}\right]\,
    \label{sigma_0}
\end{equation}
where $\Sigma_0 = 2.5 \times 10^{-6}\,\msun/$(au)$^2$, and $R_{0} = 100\,\mathrm{au}$. The initial disc mass is $M_{\rm disc}\approx 0.35\,\msun$, and the initial temperature  is given by $T(R) = 25 \,(100$~au$/R)^{1/2}$ K, which results in the minimum $Q$-parameter of $\sim 3$. The disc is irradiated by the central star, so that radiative cooling flux is given by $\sigma_B (T_{\rm eff}^4 - T_{\rm bg}^4)$, where $T_{\rm eff}$ is the disc effective temperature, and $T_{\rm bg}=20 (100$~au$/R)^{1/2}$ K.

\begin{figure}
    \centering
    \includegraphics[width=0.5\textwidth]{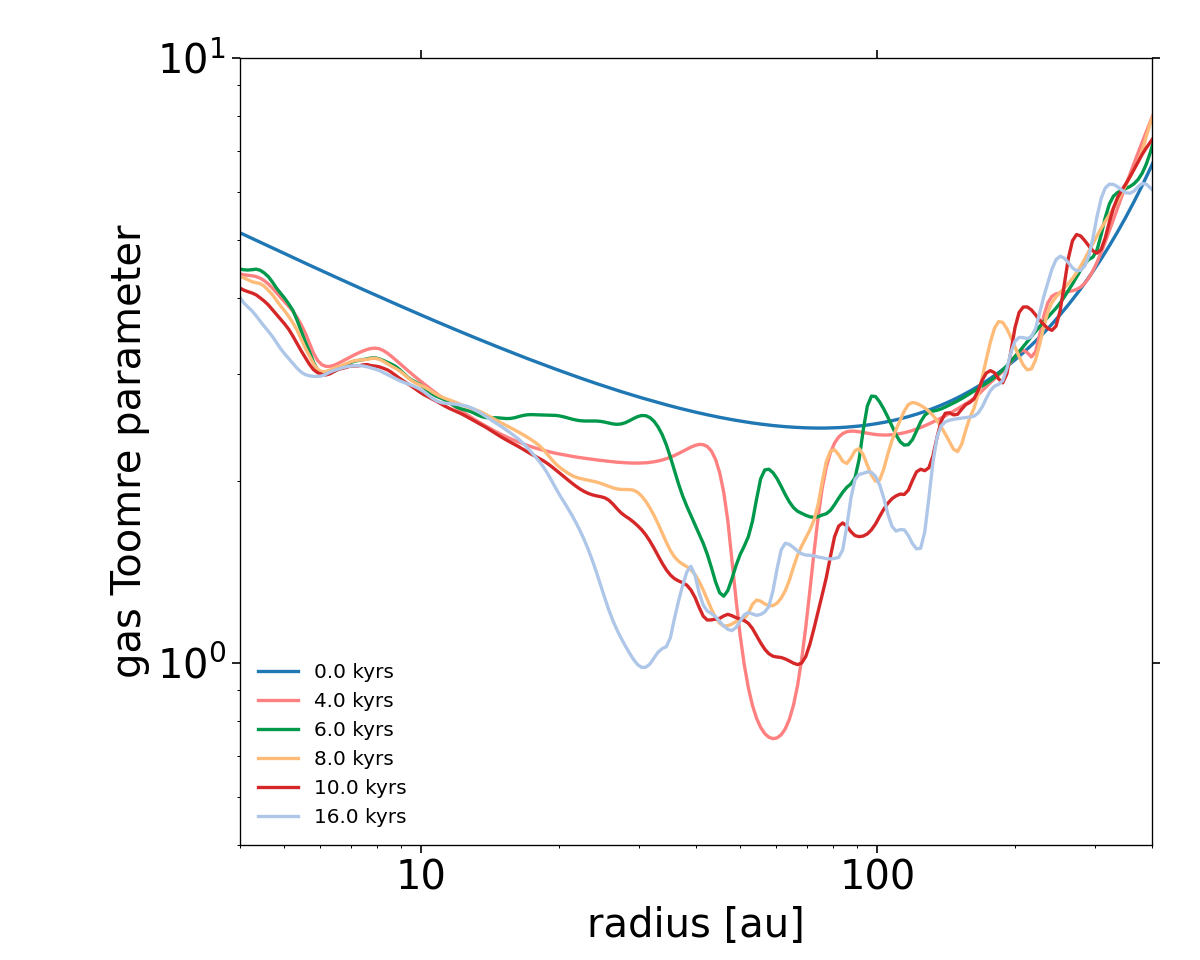}
    \caption{The azimuthally averaged Toomre $Q$ parameter for the relaxation simulation.}
    \label{fig:fargo_toomre_1D}
\end{figure}

To gently push the disc closer to smaller $Q$, following \cite{ZhuEtal12a}, we deposit additional mass into a narrow ring with radius $R_{\rm dep}=60$~au at the rate $\dot M_{\rm dep} = 10^{-5}$\MSunPerYear. We find that with these parameters, the disc becomes self-gravitating and exhibits strong spiral density arms, but does not fragment, gently self-regulating by both cooling and mass transfer \citep[][]{Longarini_25_Infall_GI}. Fig. \ref{fig:fargo_toomre_1D} shows the azimuthally averaged disc Toomre $Q$ profile for the disc. Note the formation of the $Q<1$ ring at $R=R_{\rm dep}=60$~au at $t=4$~kyr. The ring is then washed out by the mass transfer via spiral density arms, this ensures $Q \gtrsim 1$ everywhere at later times. \cite{Crida_25_reflex_instability} recently showed development of a ``reflex instability" due to ``Indirect Acceleration" terms arising from acceleration of the star due to disc-star gravitational interactions. We do not include these terms during the disc relaxation procedure, to arrive in a well behaved initial conditions; however, once the planets are injected into the disc, all of the indirect terms (from the disc and the similar terms due to the secondary and the planets) are enabled.

Similar to \texttt{PHANTOM} simulations (\S \ref{sec:phantom}), we insert a number of low mass planets and a secondary ``seed", all in circular orbits around the primary star. Once we inject the planets into the disc, we set the external mass deposition rate, $\dot M_{\rm dep}$, to zero. We ran $O(100)$ simulations varying in the exact moment when the planets are injected into the relaxed disc, the initial mass of the secondary, its accretion parameter (explained below), and injection radii of the secondary and the planets. The statistics of these simulations will be reported in a future publication. Qualitatively, the outcome of these simulations is nearly always the same: the fraction of planets ejected is large, $F_{\rm ej}\gtrsim 0.5$, as long as the secondary grows in mass to at least tens of $\mj$, and is able to migrate a significant distance towards the primary. Here we present just one simulation exemplifying this ``typical" outcome.

\begin{table}
    \centering
    \begin{threeparttable}
    \caption{Initial and final configuration of objects in the 2D simulation presented in Figs. \ref{fig:disc-2D}--\ref{fig:pla-2D}. }
    \label{tab:pla_par}
    \begin{tabular}{lcccc}
        \hline
        \textbf{Name} & \textbf{$a_0\,[\mathrm{au}]$} & \textbf{$m_0\,[\mj]$} & \textbf{$a^\text{b}_\mathrm{fin}\, [\mathrm{au}]$} & \textbf{$v^\text{c}_\infty\, [\mathrm{km}/\mathrm{s}]$} \\
        \hline
        Secondary\textsuperscript{a} & 100  & $4$ & 40 & - \\
        P1                    & 200  & $0.03$  & 500 & - \\
        P2                   & 140  & $0.01$  & 150 & - \\
        P3                    & 75   & $0.003$  & - & 1.06 \\
        P4                    & 50   & $0.015$  & - & 3.81 \\
        P5                    & 35   & $10^{-4}$  & - & 0.80 \\
        P6                    & 25   & $0.006$  & 3 & - \\
        P7                    & 15   & $0.002$  & - & 5.08 \\
        P8                    & 10   & $0.02$  & - & 6.10 \\
        P9                    & 6    & $3 \times 10^{-4}$  & 6 & - \\
        \hline
    \end{tabular}
    \begin{tablenotes}
        \small
        \item \textsuperscript{a} The secondary is allowed to accrete gas, and reaches mass of $\sim 0.1 \msun$ by the end of the simulation (cf. Fig. \ref{fig:t-m-r}).
        \item \textsuperscript{b} Final semi-major axis of surviving bound planets.
        \item \textsuperscript{c} Velocity of ejected planets at infinity.
    \end{tablenotes}
    \end{threeparttable}
\end{table}

In this simulation, 9 planets and the secondary are  initialized with masses and orbital radii listed in Table~\ref{tab:pla_par}. Note that both radial locations and masses of the planets are chosen rather arbitrary, subject only to covering a physically reasonable range in both radius and planet masses in the three different scenarios for small planet formation by disc fragmentation (\S \ref{sec:preliminaries})\footnote{We remind the reader that the range of planet masses possible in the Tidal Downsizing scenario, for example, is from sub-Earth to super-Jovian masses at any radius interior to where the disc has fragmented, see \cite{NayakshinFletcher15,Nayakshin16a}.}. Of course, in the future, it will be desirable to constrain planet locations and masses from simulations that form the planets directly.

All of the objects injected into the disc evolve under gravitational interactions with both the disc and with each other, but only the secondary is allowed to accrete gas  following the prescription of \cite{Kley_99_planet_accretion}, with slight modifications. In the prescription, each grid cell inside an accretion region inside the secondary's Hill radius, $R_{\rm Hs} = R_{\rm s} (M_{\rm s}/3M_*)^{1/3}$, loses mass into the secondary. The amount of mass lost by a cell with mass $m_{\rm cell}$ during timestep $\Delta t$ is $\Delta m_{\rm cell} = m_{\rm cell} (1 - \delta_{\rm acc})$, where $\delta_{\rm acc} = f_{\rm acc} g(r)\Omega_K \Delta t$. Here $\Omega_s$ is the Keplerian angular frequency at the secondary location, $g(r)$ is a function of distance $r$ from the planet \citep[see][]{Kley_99_planet_accretion}, and $0\leq f_{\rm acc} \leq 1$, a free parameter (here set to 0.1). Additionally, $\delta_{\rm acc}$ is capped at 0.25. In variance with \cite{Kley_99_planet_accretion} we reduce the accretion radius to $R_{\rm acc} = (3/8) R_{\rm Hs}$. The angular momentum of gas accreted by the secondary at a given time-step is added to that of the secondary. This gas accretion prescription ensures that only the gas that entered the Hill sphere of the planet and is able to cool sufficiently rapidly is accreted onto the planet \citep[e.g., see][]{Bate03,ZhuEtal12a}. While this prescription is not a first principle treatment of gas accretion onto a massive planet \citep[e.g.,][]{AyliffeBate09,AyliffeBate09b}, we emphasize that these details do not affect our main conclusions on the efficacy of FFP formation via secondary star growth and migration. While the exact evolution of the secondary, e.g., its final mass, binary separation and eccentricity, depend strongly on the parameter $f_{\rm acc}$, the fraction of ejected planets is usually similarly high whether $f_{\rm acc}=0$ of $f_{\rm acc}=1$, as long as the secondary mass is significant and its orbit evolves strongly during the simulations.

\begin{figure*}
\includegraphics[width=0.33\textwidth]{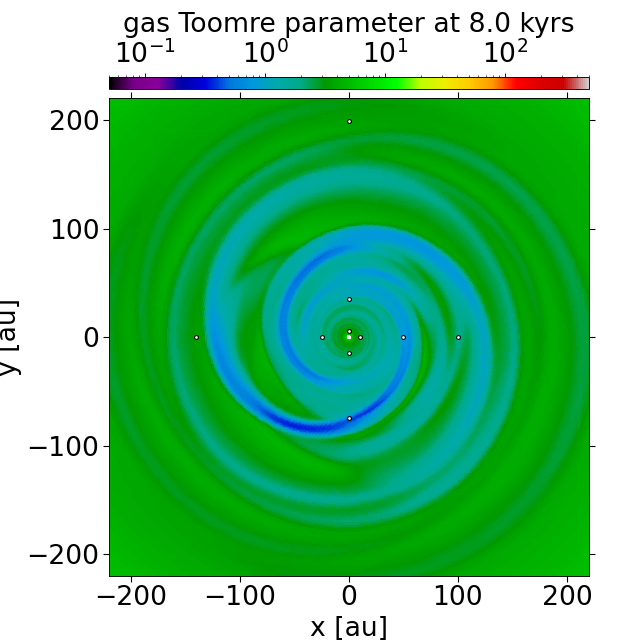}
\includegraphics[width=0.33\textwidth]{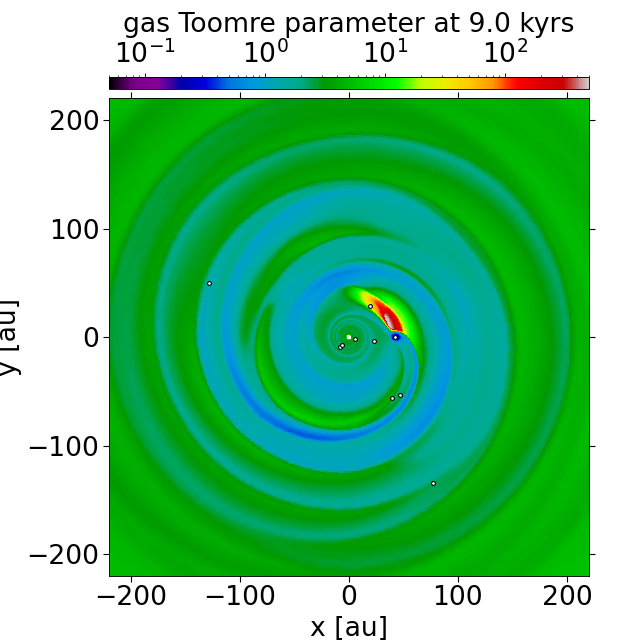}
\includegraphics[width=0.33\textwidth]{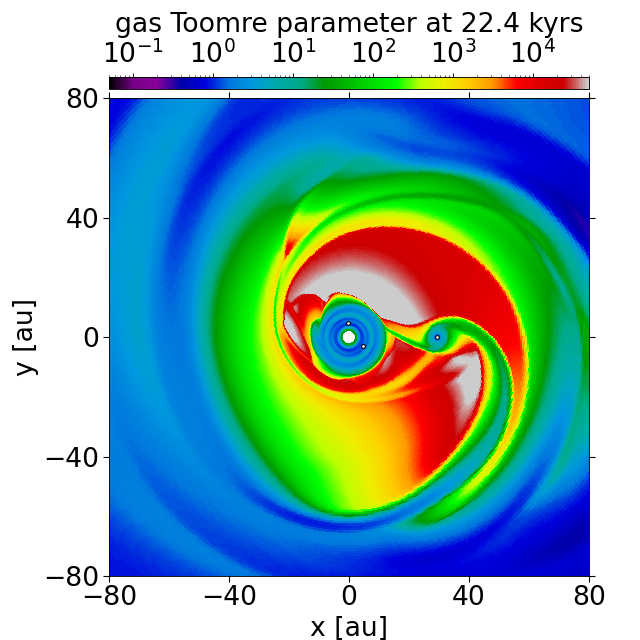}
\caption{Evolution of the disc's Toomre parameter field in the \texttt{FARGO-ADSG} simulation. White dots with black edges represent the inserted objects, while the central white region marks the inner boundary of the simulation domain. {\bf Left}: Initial state of the disc at the time of object insertion. {\bf Middle}: Formation of a clump around the secondary. {\bf Right}: Zoom-in view with extended colorbar, highlighting the deep non-axisymmetric gap opened by the secondary, and the development of spiral structures. }
\label{fig:disc-2D}
\end{figure*}

\begin{figure}
    \centering
    \includegraphics[width=0.5\textwidth]{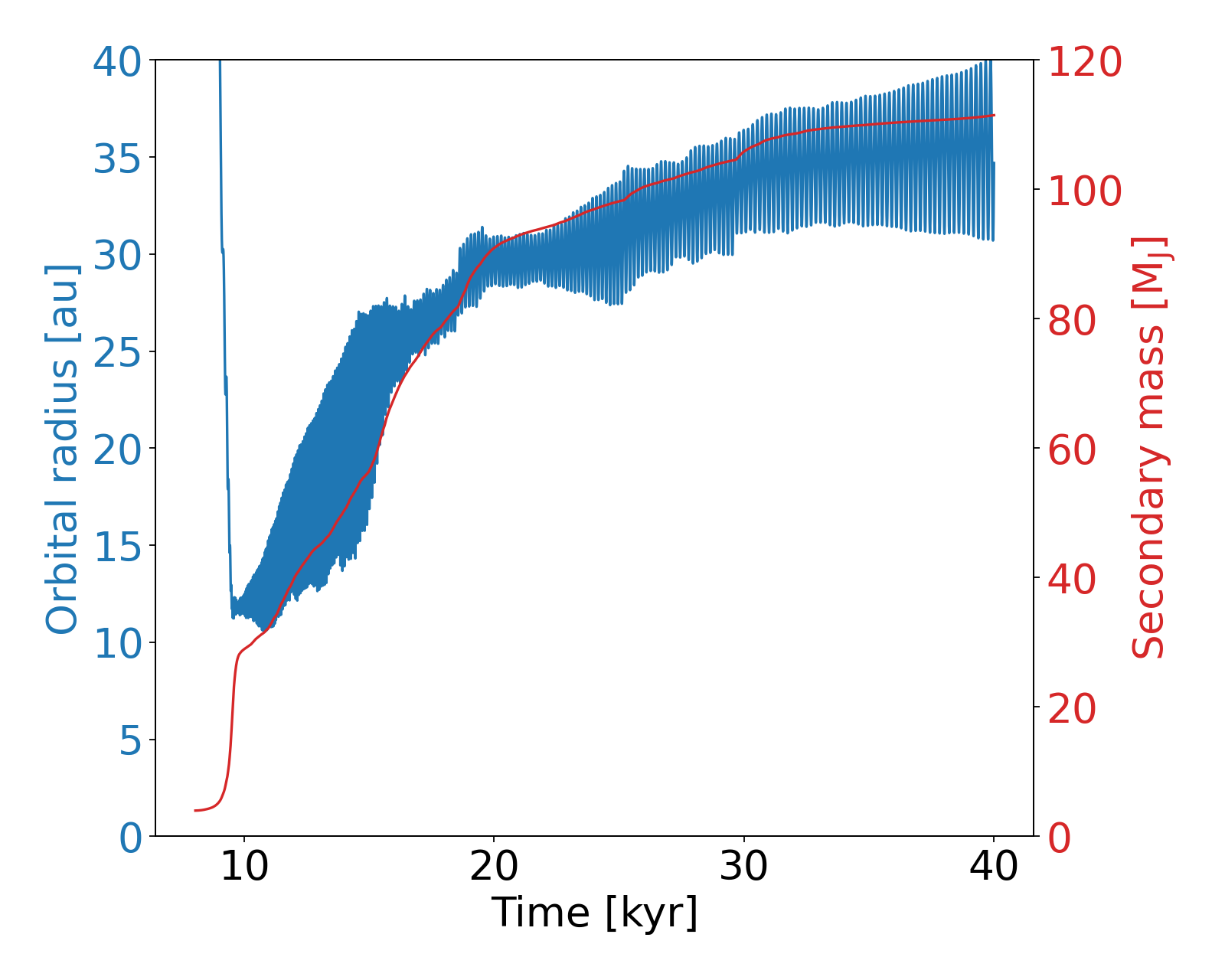}
    \caption{Evolution of the secondary's orbital radius (blue curve, vertical axis on the left) and mass (red, right axis) versus time.}
    \label{fig:t-m-r}
\end{figure}

Figure~\ref{fig:disc-2D} illustrates the evolution of the Toomre $Q$ parameter in the disc, with planets and the secondary shown as small symbols on top of the disc. Note changes in the box size and the colour bar. The calculations are performed in the frame corotating with the secondary star, which is therefore always on the $y=0$ axis in Fig. \ref{fig:disc-2D}. The secondary migrates inward rapidly, opening a deep gap in the inner disc. The radius and mass evolution of the secondary are plotted in Fig. \ref{fig:t-m-r}. The initial plunge of the secondary from 100 to 10 au only takes $\sim 2$ kyr. This is consistent with results of \cite{BaruteauEtal11}, given that our disc is about twice more massive than theirs, and $t_{\rm mig}\propto M_{\rm d}^{-1}$ (eq. \ref{tmig}). Upon reaching the inner disc, the secondary stalls since its mass is comparable at that point with the local mass of the disc there \citep[cf.][we note in passing that low mass non-accreting secondaries, $M_{\rm s} \lesssim $ a few $\mj$, do not stall but migrate all the way to the inner boundary of our computational domain]{MalikEtal15}. The secondary then opens a deep  gap in the disc not only due to its torques onto the disc but also by gas accretion. Fig. \ref{fig:t-m-r} shows an interesting positive correlation between the secondary mass and its semi-major axis about the primary star. The secondary's orbit expands with time from $t \gtrsim 10$~kyr because it accretes gas from the circumbinary disc at that point, which has a larger specific angular momentum than the secondary. By $t=40$ kyr, the secondary becomes a low mass star.

\begin{figure*}
    \centering
    \includegraphics[width=0.4\textwidth]{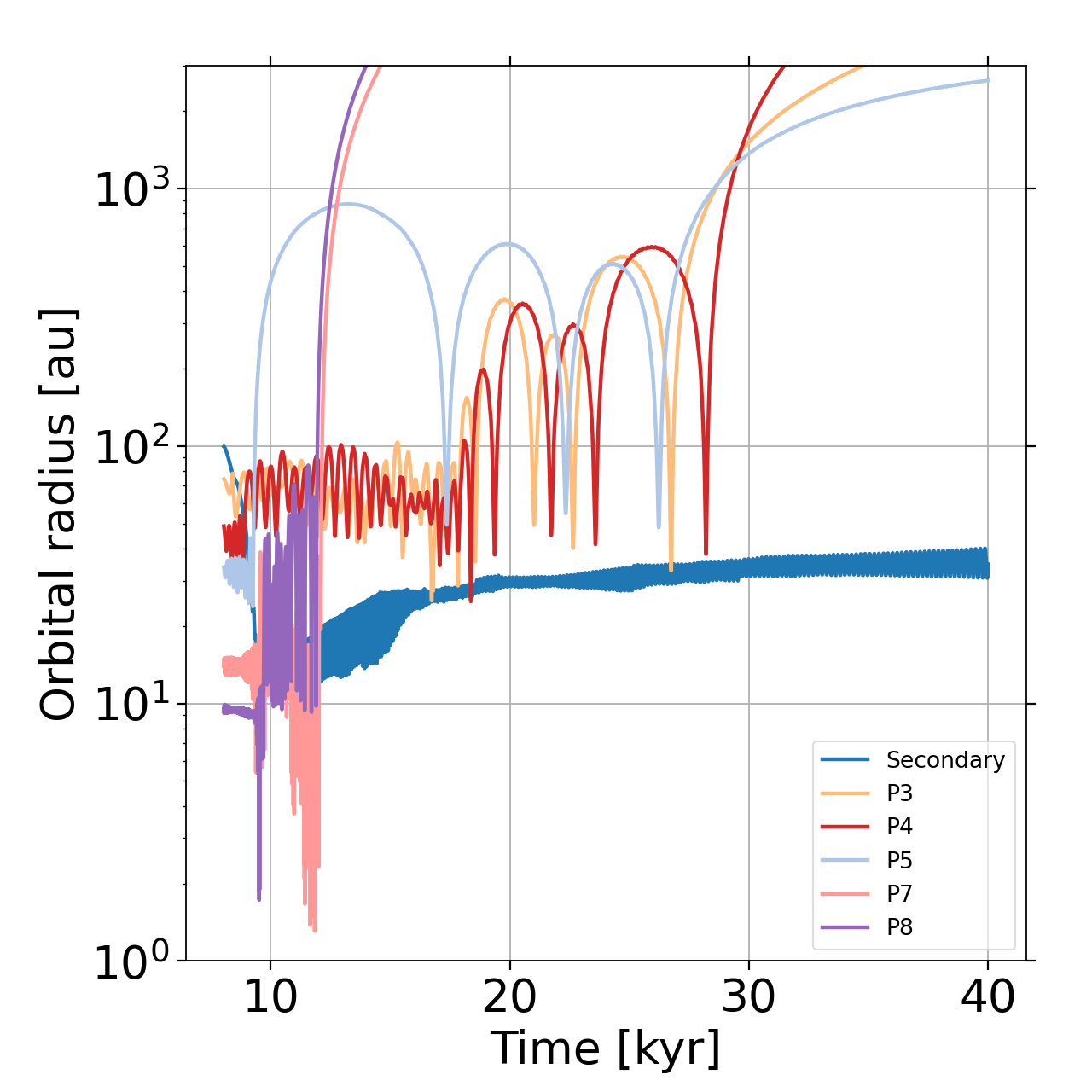}
    \includegraphics[width=0.4\textwidth]{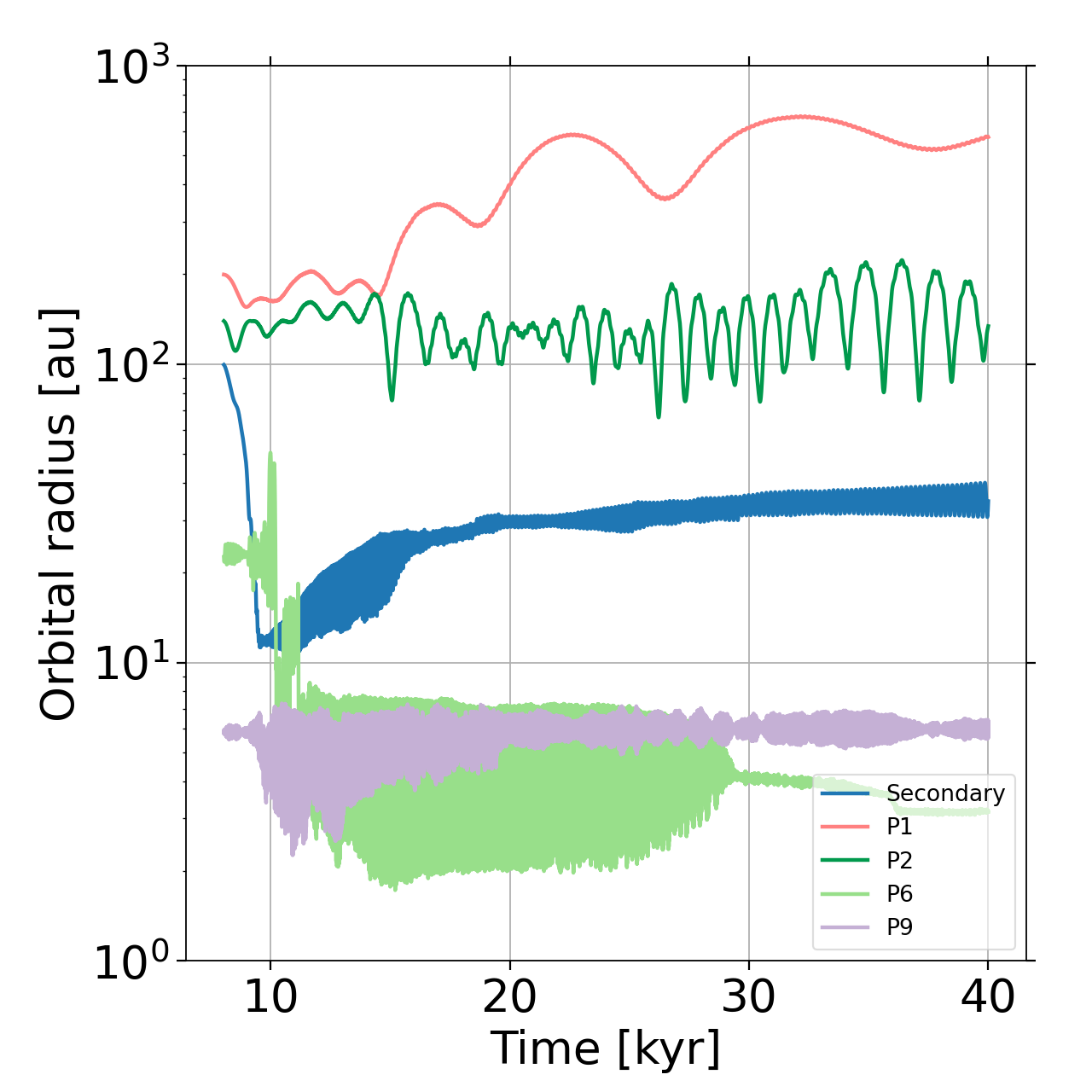}

    \includegraphics[width=0.4\textwidth]{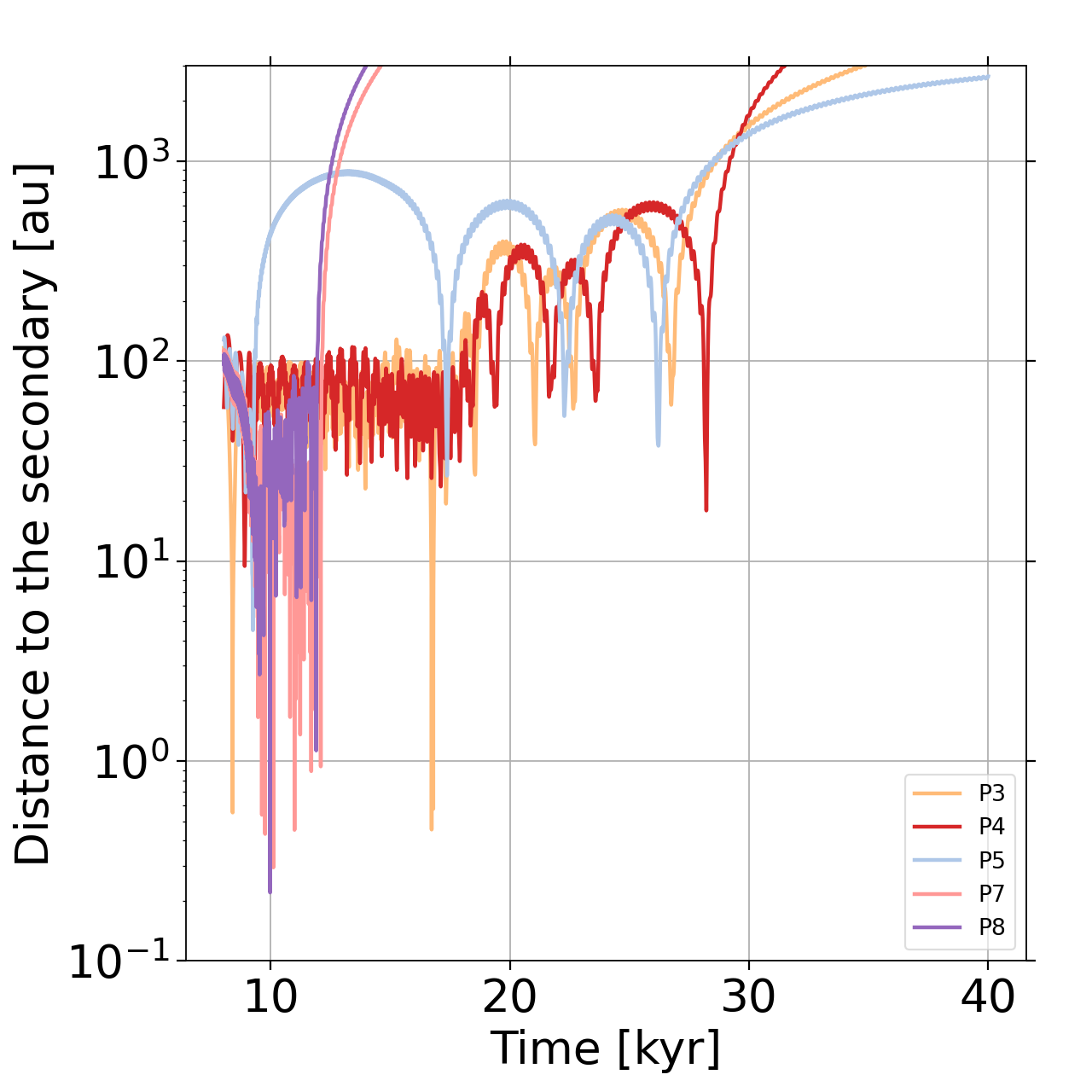}
    \includegraphics[width=0.4\textwidth]{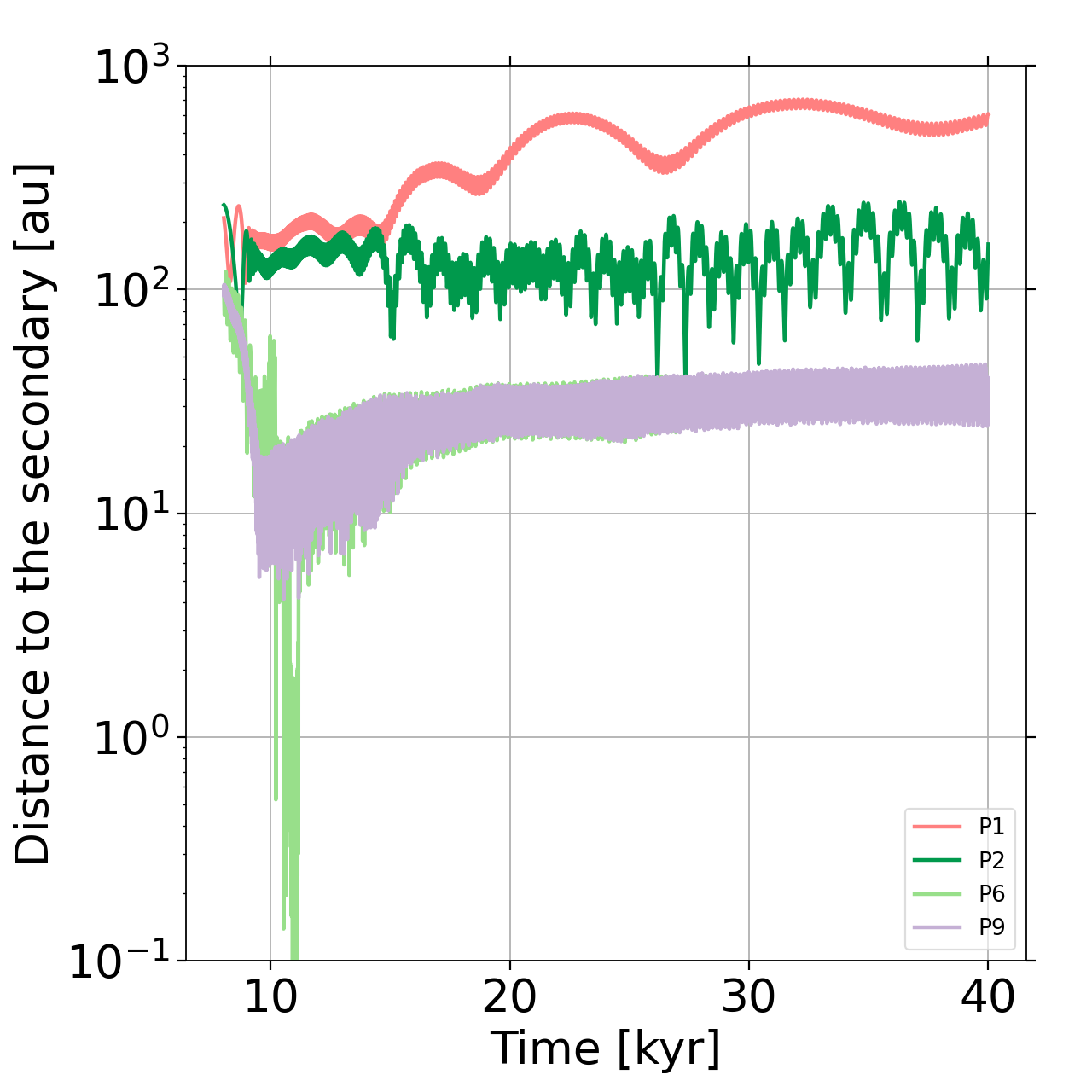}
    
    \caption{Orbital radius (top panels) and distance to the secondary (bottom panels) for two groups of planets in the \textsc{FARGO-ADSG} simulations: those ejected from the system (left panels) and those remaining bound (right panels). The outermost planets and the innermost planet survive secondary formation; planet P6's orbit shrinks strongly after interaction with the secondary.}
    \label{fig:pla-2D}
\end{figure*}

Since the secondary sweeps through the tens of au disc region {\em twice}, a large range of planets are susceptible to ejections. Fig.~\ref{fig:pla-2D} shows the orbital radius and distance to the secondary of  ejected (left panels) and surviving (right panels) planets. Table \ref{tab:pla_par} gives the velocities of the ejected planets and semi-major axis of the survivors, respectively. We see that, by large, the planets with initial values of $a_0$ in the disc region swept by the secondary are the most vulnerable to ejections, whereas the planets outside this radial range have better chances of survival. Yet there are exceptions. For instance, P6 survives by being scattered on a smaller orbit and an ensuing circularisation of its orbit by the disc. The two planets on initial orbits larger than the secondary's orbit survive since they did not have a close encounter with it. 

In line with what found by \cite{Calovic_25_FFP-1} in 3D SPH simulations, and in our \texttt{REBOUND} simulations, \texttt{FARGO-ADSG} simulations result in velocity of ejectees at infinity being generally low, i.e., between $\lesssim 1$ and a few km/s (cf. Table \ref{tab:pla_par}). However, there are some interesting differences that can be traced to the disc becoming  eccentric at later times.  For example, due to non-circular disc torques, the outermost planet is scattered onto a much wider orbit (see the right panel in Fig. \ref{fig:pla-2D}).

\subsection{Planet ejection efficiency versus mass of the secondary}\label{sec:fargo_statistics}



 
 To test sensitivity of our main results to the mass of the secondary object, we run additional simulations in which gas accretion on the secondary is disabled, and the secondary companion massses are fixed at  0.5, 1, 2, 4, 8, 16, 32, or 64~$M_{\mathrm{J}}$. The parameters of the low mass nine planets injected into the disc are identical to those in the fiducial model (see Table~\ref{tab:pla_par}). For these runs, mass injection into the disc was terminated, and the disc was relaxed for an additional time, which resulted in a smoother Toomre parameter profile. For each secondary mass, six simulations were carried out, with two values of the initial orbital radius for the secondary (100 and 125~au), and three planet/secondary injection times (26, 27 and 28 kyr). 

\begin{figure*}
\centering
\begin{subfigure}[t]{0.3\linewidth}
\centering
\includegraphics[width=\linewidth]{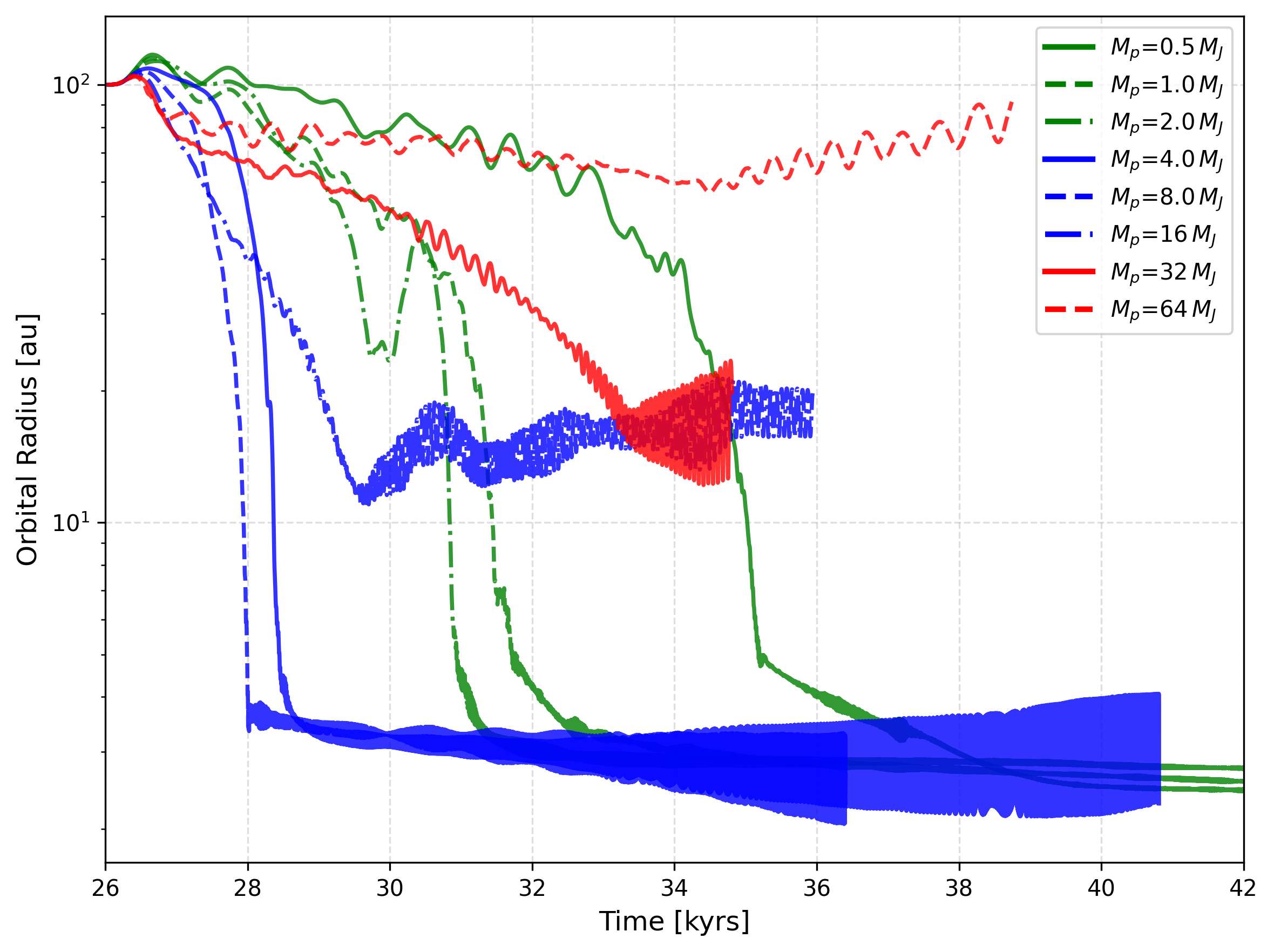}
\caption{Orbital evolution of secondary objects injected at $R_0 = 100$~au and time $t_{\mathrm{inj}} = 26$~kyr.}
\label{fig:sec-motion}
\end{subfigure}
\hfill
\begin{subfigure}[t]{0.3\linewidth}
\centering
\includegraphics[width=\linewidth]{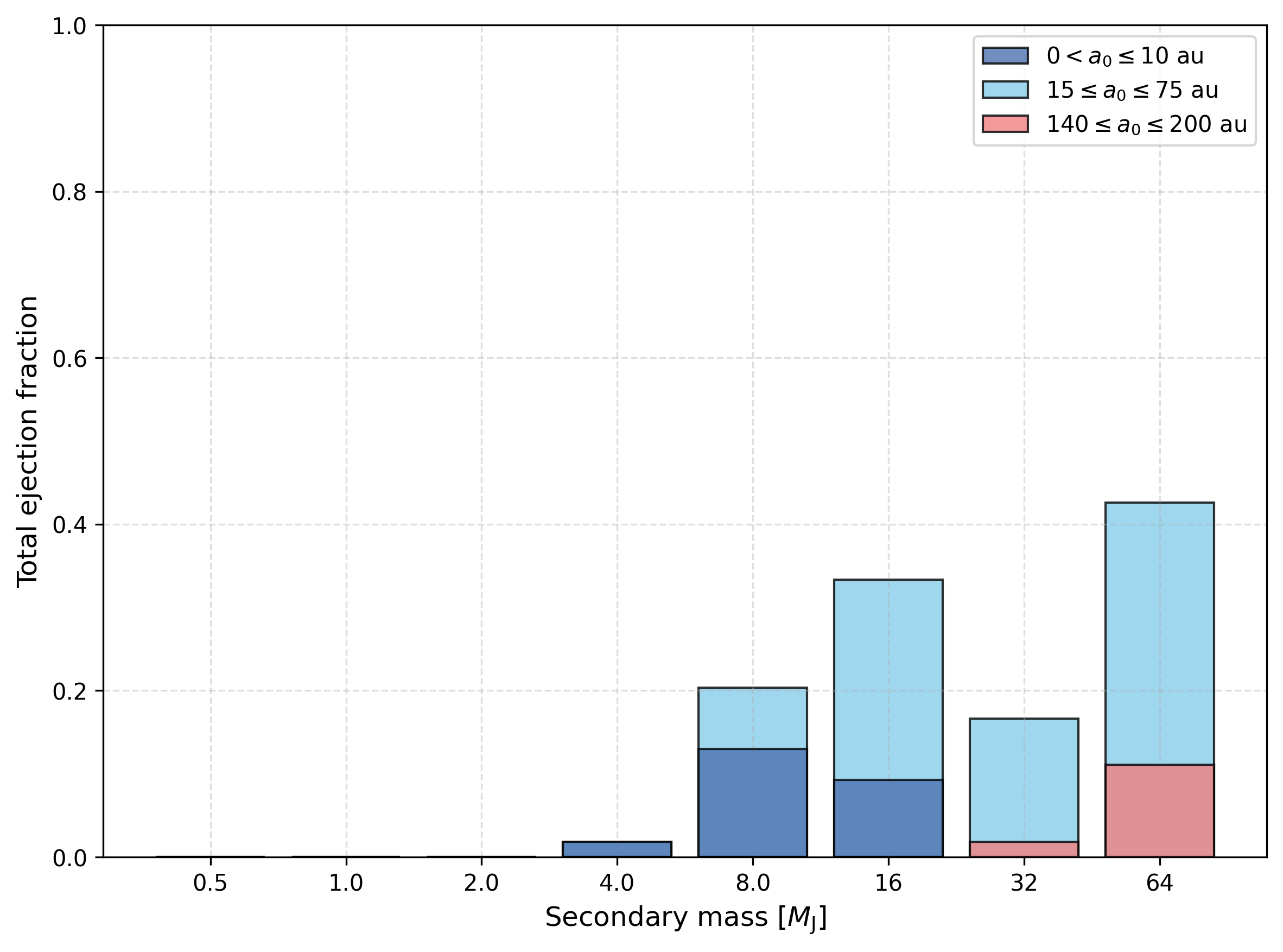}
\caption{Total ejection fraction of planets versus secondary mass. Colours mark three ranges of planet initial orbital radii, $a_0$.}
\label{fig:eject_tot}
\end{subfigure}
\hfill
\begin{subfigure}[t]{0.3\linewidth}
\centering
\includegraphics[width=\linewidth]{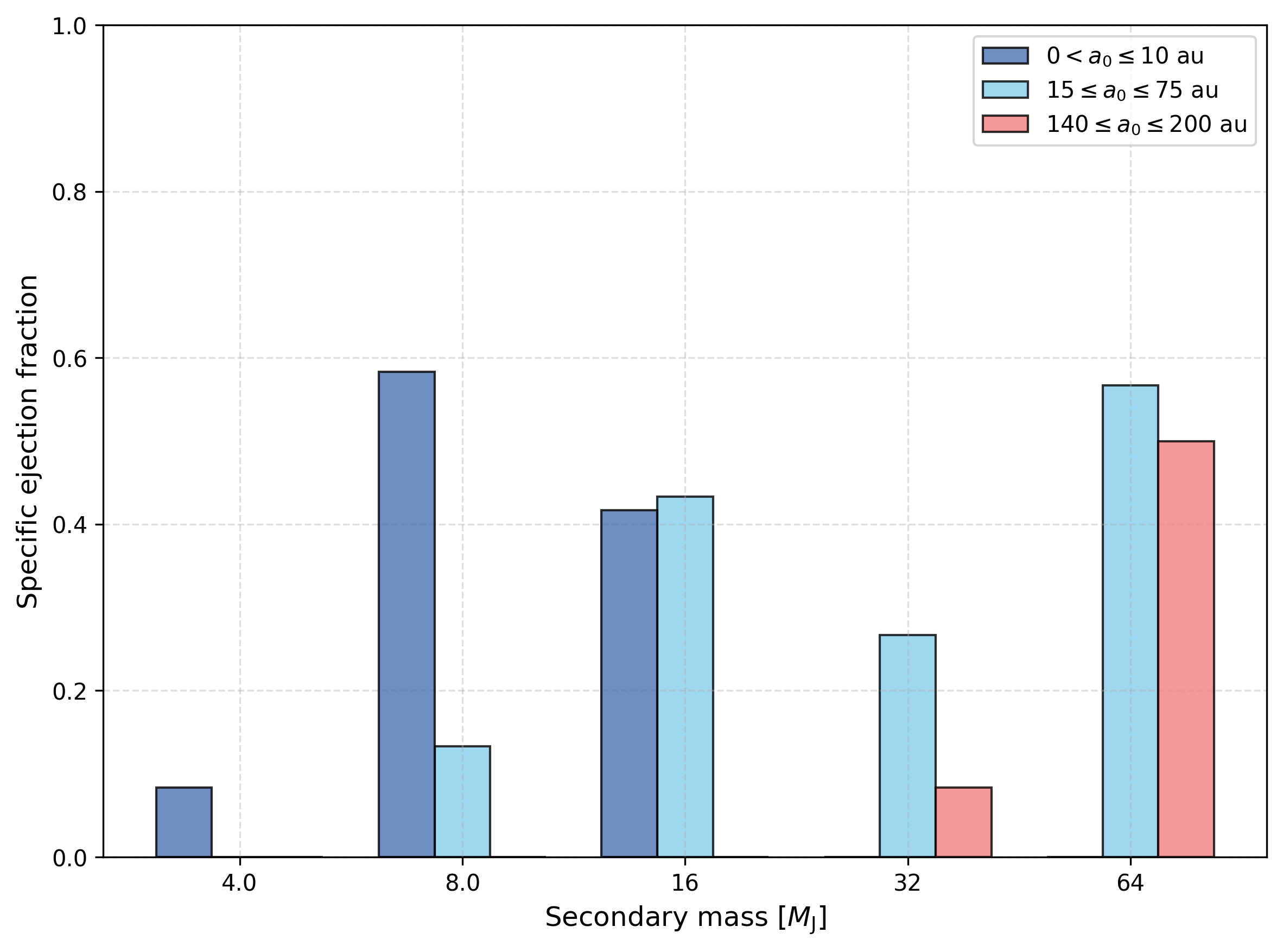}
\caption{Specific ejection fraction for planets within radial ranges specified in the legend.}
\label{fig:eject_sep}
\end{subfigure}
\caption{
    Results of a suite of simulations described in \S \ref{sec:fargo_statistics}.}
\label{fig:eject_fractions}
\end{figure*}

Figure~\ref{fig:sec-motion} shows the secondary migration tracks for runs with injection time of 26~kyr and the secondary starting off  at 100~au. All lower mass objects (0.5–8~$M_{\mathrm{J}}$) undergo rapid inward migration halted only very close to the inner boundary. This is a limitation of our simulations; it is likely that in simulations with a smaller inner grid boundary, the rapid migration of low mass secondaries would continue to smaller radii. Higher-mass companions ($\geq$16~$M_{\mathrm{J}}$), on the other hand, do not reach the innermost regions of our computational domain, stalling and then migrating outwards somewhat (similar to the secondary in the fiducial run, Fig. \ref{fig:pla-2D}). High mass companion stalling is expected based on Fig.~\ref{fig:theory_prelims}, which shows that objects with $M_{\mathrm{p}} \gtrsim  16~M_{\mathrm{J}}$ are sufficiently massive to open a deep gap in the disc, at which point they switch into a much slower type-II migration regime \citep[e.g.,][]{MalikEtal15}.

These secondary object migration patterns affect low mass planet ejection strongly. 
Figure~\ref{fig:eject_tot} shows the total planet ejection fraction versus the secondary object mass. This is defined as the total number of planets ejected divided by 54 (the total number of planets in all 6 runs). Colours in Fig.~\ref{fig:eject_tot} show the contribution of planets  within a specific range of initial orbital radii to that total. We observe that planet ejections occur only when the secondary mass exceeds 4~$M_{\mathrm{J}}$. This is in contrast to results of many previous studies that considered two or more gas giant planets in tightly packed configurations \citep[e.g.,][]{Weidenschilling_96_FFPs,Rasio_96_FFPs,juric2008,veras-12-FFPs}. In our problem setup, we consider only one secondary object embedded in a massive protoplanetary disc. The low mass planets have relatively large initial orbital separations, and if the secondary sweeps through that radial range rapidly, then only the sufficiently massive of them provide strong enough kicks to eject low mass planets. Fig.~\ref{fig:eject_tot} agrees qualitatively with the idealised analytical model in \S \ref{sec:pincer}, which shows that the secondary mass should be greater than a few to 10 \% of $M_*$ for efficient ejections in the GI scenario. This is also supported by the results of \cite{Calovic_25_FFP-1}. On the other hand, it is possible that gas giant planets would eject low mass brethren on longer time scales that we do not explore here.

Fig.~\ref{fig:eject_sep} presents the specific ejection fraction, defined as the fraction of planets ejected only from within a range of initial radii. The figure demonstrates that to eject a planet, the secondary must be not only massive but also spend enough time on a similarly sized orbit \citep[as expected based on the][criteria]{Holman_Wiegert_99}. For instance, planets P8 and P9 (initial orbital radii $a_0 = 6$ and 10~au) are ejected by secondaries with masses between 4 and 16~$M_{\mathrm{J}}$ only; higher mass secondaries stall on orbits too wide to affect these two planets. For the same reason, secondary objects with $M_{\mathrm{p}} = 8~M_{\mathrm{J}}$ eject more than half of the innermost planets ($a_0 \leq 10$~au) but only $\sim$13\% of planets in the range $15 \leq a_0 \leq 75$~au. Finally, the outermost planets (P1 and P2; 140–200~au) are ejected by the most massive companions (32–64~$M_{\mathrm{J}}$), with the ejection fraction exceeding 50\% for $M_{\mathrm{p}} = 64~M_{\mathrm{J}}$.

Comparing these results with the fiducial run, we see that secondary objects growing in mass from the planetary to the stellar mass regime may be most efficient in ejecting low mass planets. This is because such objects are able to cover a broad range in orbital separations before opening the gap, they also tend to migrate in and then out, and they can become very massive eventually (Fig. \ref{fig:t-m-r}).


\section{Discussion}\label{sec:discussion}

\subsection{The FFP challenge to planet formation}

The surprisingly abundant population of microlensing FFPs pose  challenges to the Core Accretion scenarios of planet formation. In single star system, there are too few massive planets to eject enough FFPs into the field. In binaries, planet formation via CA is only possible in close separation, $a_{\rm b} \lesssim 1$~au, or very wide systems, $a_{\rm b}\gtrsim 100$~au. The observed rarity of planets in binaries in this separation range \citep{Moe_21_planets_in_binaries,Thebault_25_planets_in_binaries} is often interpreted as evidence for difficulties of their formation in a typical binary system \citep[e.g.,][]{Thebault_15_Stype_binaries}.  If this were so, then it is hard to understand how binary systems can ever eject enough planets (see \S \ref{sec:CA_FFP}) to produce an FFP population rivalling or even exceeding the bound population of planets. Indeed, close binaries are needed for an efficient FFP ejection \citep{Coleman-24-FFP-simulations,Coleman-24-FFPs}, but these are rare, and on the top of that, the binary fraction  for M-dwarf stars is also relatively small \citep[$\sim 0.25$;][]{Offner_23_binaries_review}.

\subsection{General framework for disc fragmentation FFPs}\label{sec:discussion_framework}

We proposed that most Free Floating Planets form by disc fragmentation (GI) in very young (Class 0/I) binary systems. GI has already been invoked in previous literature to account for gas giant planets in M-dwarfs and Brown Dwarfs in FGK stars (see Introduction). 

Disc fragmentation is only possible at the outer fringes \citep[$R\gtrsim $~tens of au;][]{Rafikov05,Clarke09} of very massive protoplanetary discs. In our scenario, single star discs either never fragment because they are too small or fragment on planet/BD mass objects. Planet migration brings some of the planets and BDs all the way into the inner disc \citep[e.g.,][]{BoleyEtal10,BaruteauEtal11}. A fraction of these objects remains stranded on wide orbits \citep[e.g.][]{ForganRice13b,NayakshinFletcher15,MullerEtal18,HumphriesEtal19,Schib_25_dipsy1,Schib_25_dipsy2}. Either way, most of these GI-formed planets/BDs remain bound to the their single star hosts.

We propose that a typical binary system \citep[separation $a_{\rm b}\sim$ tens of au;][]{RaghavanEtal10} starts out from a single star configuration. The disc is assumed to fragment and make planets at tens to $\sim 100$ au. As it continues to grow due to mass deposition from the collapsing molecular cloud \citep[e.g.,][]{VB06,VB10,VB15}, most massive fragments are expected to form and grow by gas accretion at widest separations, $R\gtrsim 100$~au (\S\S \ref{sec:initial_fragments} \& \S \ref{sec:accretion_prelim}). We conjecture that the seed of the secondary star is a planet oligarch that formed in these outer disc fringes.  Due to its larger initial mass and favourable location, where radiative cooling is more efficient, the seed grows and also migrates inward (\S \ref{sec:planets_or_binary}) on time scales of just a few orbits (i.e., a few thousand years). Numerical simulations further show that growth in mass takes a runaway character when the secondary's migration stalls. The final mass of the secondary is likely to be a good fraction of disc mass, resulting in a binary with mass ratio $q = 0.1-0.3$, and possibly larger, if the mass infall from the parent molecular cloud continues. 

The resulting reconfiguration of the system from a ``single star with some planets" state into a ``young growing binary system with some planets in the midst" is very rapid and violent. Shrinking of binary separation leads to abundant planet-secondary star scatterings that give stochastic velocity kicks to the planets via the ``gravity assist" effect. Analytical arguments (\S \ref{sec:pincer}) show that secondaries with mass ratio $q\gtrsim 0.05$ are likely to eject most of the planets that they encounter out of the system.

A central element in our scenario is the assumption that disc fragmentation can produce objects as diverse in mass/composition as planetesimals to Earth mass planets to gas giants/BDs to secondary stars. These suggestions are not new \citep[e.g.,][proposed that the Solar System planets and stellar binaries both form via GI]{Kuiper51}. While contemporary astrophysical literature is full of statements ``GI does not make sub-giant planets" (or similar), in fact, three different disc fragmentation scenarios for forming small ($M_{\rm p} < $ Saturn mass) planets exist (cf. \S \ref{sec:accretion_prelim} \& \S \ref{sec:MHD_or_dust_prelim}). All three of these schemes favour small planet birth locations of tens to $R\lesssim 100$~au. 


\subsection{Observational tests}\label{sec:discussion_future_obs}

\subsubsection{FFP numbers}\label{sec:discussion_numbers}

There are  roughly $\sim 10$ Earths and $\sim 2$ Neptune mass rogue planets per M-dwarf star \citep{sumi2023,Mroz-23-Microlensing-planets-review}. Low mass stars feature just $\sim 0.06$ bound gas giants per star, so it is not likely that they could eject enough FFPs. M-dwarf binary fraction is also low, $\sim 0.25$, but larger than that of gas giant planets, implying that binary stars are more likely culprits. However, whatever mechanism forms planets in binaries, it would have to be probably more efficient than it is in single stars (even if the FFP population is over-estimated by a factor of a few). Placing such a requirement on any of the CA scenarios appears counter-intuitive as all of them are expected to be less (not more) efficient in binaries. 

On the other hand, there may be physical reasons for disc fragmentation hatching more planets in binaries than in discs of single stars. To make a secondary star, an especially massive disc is needed; a more massive disc is likely to fragment on more planets. Secondary stars can also trigger fragmentation when none is expected otherwise \citep{Meru15,Cadman_22_triggered_fragm_binary}. Furthermore, in our scenario (a) there is no ``binary system penalty" associated with forming a planet in a binary system because the secondary star begins, as any star, from an object with a modest mass of a few to 10 Jupiter masses \citep[a First Core;][]{Larson69}; (b) Our scenario works very well for a garden variety binary with {\em final} separation of $a_{\rm b}\sim 40$~au, whereas for CA only the (rare) close binaries can be efficient planet ejectors \citep[e.g.,][]{Coleman-24-FFPs}.

\subsubsection{Masss function}\label{sec:mass_function}

CA predicts few FFPs below $M_{\rm p}\sim 10 \mearth$ \citep[see Figs. 2-4 in][]{Coleman-24-FFPs} because lower mass planets are destroyed by collisions and also do not migrate inward to be ejected rapidly enough. In contrast, in our scenario, ejection of even very low mass planets, e.g., of Lunar mass and even comet mass objects, is similarly efficient. The crucial difference between CA and GI scenarios for FFP ejection is what ``moves".
In the former, a would-be FFP object needs to migrate to the secondary star to receive its kick; this does not occur rapidly enough for low mass objects \citep{Coleman-23-planet-in-CB-systems,Coleman-24-FFP-simulations} since they migrate in the type I regime, where $t_{\rm mig}\propto M_{\rm p}^{-1}$ \citep{BaruteauEtal14a}. In our scenario, it is the secondary star that catches up with the small objects, and all objects less massive than $\sim 1\mj$ are effectively stationary and also low mass (test) particles compared to the secondary, so all are ejected equally efficiently. Our scenario hence predicts no ``ejection deficit" for FFPs below the $\sim 10\mearth$, in contrast to CA \citep[e.g., \S 3 in][]{Coleman-24-FFP-simulations}.  

Further, we expect that the FFP mass function should be less steep than the microlensing bound planet mass function. Broadly speaking,  disc fragmentation FFP mass function, $dN/d M_{\rm p}$, is a product of the planet mass function at the time of binary formation, $dN_0/d M_{\rm p}$, and the ejection probability, $p_{\rm ej}(M_{\rm p})$. As explained above, for planets migrating inward slowly, $p_{\rm ej}(M_{\rm p})\sim 0.7$, i.e., a constant. For these planets, then, $dN/d M_{\rm p}\sim dN_0/d M_{\rm p}$. If most of the bound microlensing planets are in single star systems (most M-dwarfs are not in binaries), then we can assume that $dN_0/d M_{\rm p}$ is somewhat similar to that of the bound microlensing planet population. Planets in gas giant planet mass range, however, do start to migrate inward appreciably. In our scenario, the secondary star's time of formation is offset from that of a planet formation. Across a population of binary systems and their planets, this offset is likely to be distributed from 0 to $\sim 0.1$ Myr \citep[as a typical molecular cloud collapse time scale;][]{Larson69}. The more massive is a planet, the more likely it is to migrate sufficiently deep towards the primary star to be immune to the secondary's gravitational perturbations. 

A population study is needed to detail this, but the statistics of a few dozen \texttt{FARGO-ADSG} simulations, to be reported elsewhere, supports these ideas. Further, in \cite{Calovic_25_FFP-1} we explored the same FFP formation scenario but only for planets with masses $(1-3)\mj$, and found that the $3\mj$ planets were the least likely to be ejected because the secondary star had difficulty ``catching them" (Fig. 4-5 in \cite{Calovic_25_FFP-1}).

This prediction is consistent with the  observations, within the current observational uncertainties. For the $\mu$FFP mass function, $dN/dM_{\rm p}\sim 1/M_{\rm p}^2$ (see Introduction for references). Looking at the ratio of the number of planets with masses $M_1$ and $M_2$, we have $N_1/N_2 \sim M_2/M_1$. For example, for $M_1=10 \mearth$ and $M_2= 3\mj$, we have 
\begin{equation}
    \frac{N_1}{N_2} = \frac{3 \mj}{10\mearth} \approx 100\;.
    \label{N_12_FFPs}
\end{equation}
In contrast, bound microlensing planet mass function is less steep, $dN/dM_{\rm p}\sim 1/M_{\rm p}^{1.6}$ \citep{SuzukiEtal16,SuzukiEtal18}, which yields $N_1/N_2 \sim (M_2/M_1)^{0.6}\approx 16$. More recently,  \cite{Zang_25_bound_microlensing} deduced that a sum of two broad Gaussians centred on $\sim 7.5\mearth$ and $\sim 2.5\mj$ is a better fit to the bound planet mass function that a power-law. Their mass function yields a similar mass ratio, $N_1/N_2 \sim 10$. The bound microlensing population is thus indeed more top-heavy than the FFPs, but the conclusion is only valid within $\sim 1.5 \sigma$ due to the uncertainties in the FFP mass function (P. Mroz, private communication).

\subsubsection{Particulate/debris discs and rings around FFPs}\label{sec:discussion_rings}

\cite{Calovic_25_FFP-1} argued that gas giant mass disc fragmentation FFPs may carry gas and dust discs with them, with radial size of up to $\sim 1$ to a few au. Here we focused on smaller mass planets, which can carry away discs with size of a fraction of their Hill radius, which translates to (typically) $\sim 0.1$~au. Circum-FFP gas discs would be dispersed quickly, but if the discs contained planetesimals \citep[see][]{NayakshinCha12}, then these could be a long-lived source of solid debris rings formed by fragmentation cascades, perhaps similar in appearance to the Saturn rings. The CA FFP ejection process, in contrast, is much less likely to contain such rings. CA FFPs are ejected from much deeper in the potential well of a close binary ($a_{\rm b}\lesssim 1$~au), so the ejection process is much more violent, resulting in an order of magnitude higher ejection velocities. 





\section{Conclusions}

In this paper we presented a scenario in which Free Floating Planets are flung out from very young growing binary systems. In this picture, the planets form by disc fragmentation {\em before} the secondary star forms. We then argued analytically that, as the secondary star grows in mass and migrates closer to the primary, many of the planets will be ejected out of the system via close interactions with it. We presented proof-of-concept simulations with three different codes, which confirm  this scenario's high efficiency of planet ejections. We concluded with a broad discussion of how our ideas can be examined with future observations.

\section{Aknowledgement}

The authors thank Przemek Mroz, Fabo Feng and Philippe Thebault for useful discussions. AC and HL acknowledge support by STFC PhD studentships. LZ is supported by the China Scholarship Council (CSC) in collaboration with the University of Leicester. FM acknowledges support from The Royal Society Dorothy Hodgkin Fellowship

This research used the ALICE High Performance Computing facility at the University of Leicester, and the DiRAC Data Intensive service at Leicester, operated by the University of Leicester IT Services, which forms part of the STFC DiRAC HPC Facility (\href{www.dirac.ac.uk}{www.dirac.ac.uk}). 

\section{Data availability}

The data obtained in our simulations can be made available on reasonable request to the corresponding author.



\bibliographystyle{mnras}
\bibliography{nayakshin,Luyao}

\bsp	
\label{lastpage}
\end{document}